# CPS Engineering: Gap Analysis and Perspectives

Emmanuel LEDINOT
**THALES Research & Technology France**

## Abstract

*Virtualization of computing and networking, IT-OT convergence, cybersecurity and AI-based enhancement of autonomy are significantly increasing the complexity of CPS and CPSoS. New challenges have emerged to demonstrate that these systems are safe and secure. We emphasize the role of control and emerging fields therein, like symbolic control or set-based fault-tolerant and decentralized control, to address safety. We have chosen three open verification problems we deem central in cost-effective development and certification of safety critical CPSoS. We review some promising threads of research that could lead in the long term to a scalable and powerful verification strategy. Its main components are set-based and invariant-based design, contracts, adversarial testing, algorithmic geometry of dynamics, and probabilistic estimation derived from compositional massive testing. To explore these orientations in collaborative projects, and to promote them in certification arenas, we propose to continue and upgrade an open innovation drone-based use case that originated from a collaborative research project in aeronautic certification reformation.*

# Introduction

From the 70s to the early 2010s, development of systems featuring real-time control of physics has been named “embedded system engineering” or “automation”. The embedded systems, e.g. defence systems, power plants, transportation vehicles, health care devices, etc., have reached over these decades high levels of complexity often measured by the size of embedded software (Klocs, Mlocs, etc.). Early on these embedded systems have featured networking aspects. First internally to connect the sensors, actuators, micro-controllers and mission computers, then externally by means of wireless communications. This outer-connectivity has recently blossomed with mobile devices and with the Internet of Things, also named Machine2Machine or Internet of Everything.

Why then the emergence, by 2006-2010, of the “Cyber Physical System” appellation to denote systems that seem identical to what has been developed so far in Operational Technologies (OT)? In other words, is “CPS” some rebranding intended to cope with the latent perception that system engineering in general, and embedded system engineering in particular, are mature fields that no longer deserve to be in limelight?

We start by analyzing “Cyber” of “Cyber-Physical”. It is not “Cyber” as meant by cyber of cyber-security, though cybersecurity admittedly is of prominent importance in CPS engineering. We support the opinion according to which some characteristics distinguish CPS from embedded systems and automation. We review some of them with safety and behavioral complexity in mind. Then, we propose three CPS engineering problems we think remain open for academic research and industrial practice altogether. They have always been, and still are, open problems for embedded systems and system engineering in general. However, with the consent of many, and occasionally huge, budget and schedule overruns, system engineering managed to cope with them, and in the end to deliver good quality products and services. Based on three long lasting problems we chose among many other ones, we argue that the lag of engineering w.r.t. to product and operation complexity is likely to worsen significantly with CPS, and even more so with CPSoS.

In the third section, we present an open innovation CPSoS end-to-end development use case initiated in aviation certification research by 2015-2018. We propose to continue collaborative research on an extended and refactored version, in search for solutions to the three selected open problems. They all focus on behavioral specification, design and verification. Finally, we present three groups of research work we deem have some potential to fill these three pointed gaps: set-based system engineering, geometric analysis of dynamics, and paradigm shift in safety engineering.

# 1. What are Cyber-Physical Systems?

A significant number of expert groups have addressed this question worldwide, for instance [P4C18] and [EF20] in Europe, yet without getting to consensus. We contribute some perspectives, motivated by certifiable autonomy and IT-OT convergence one side, by some feeling of under estimation on the other side.

In the late 40s, the mathematician Norbert Wiener perceived the prominent role of *feedback* to understand the behavior of natural and artificial systems. He coined the term *Cybernetics* [Wie48]. He derived it from the greek “Kubernêtiké”, whose meaning is “the art of governing”. Since then, “Control” has superseded “Cybernetics”. However, “Cyber” did not disappear whatsoever, as witnessed by “cyber-security”. The meaning of “cyber” seems to have shifted from “governing” to some synonymous of “digital”.

In this paper, we promote the primal interpretation of “Cyber”, i.e. *steering, governing,* as the prominent feature of Cyber Physical Systems. This stance is helpful to understand why CPS engineering raises new challenges compared to embedded system engineering and IT system engineering. Control theory, fault-tolerant control, networked control, compositional control, symbolic control, and control-oriented safety are examples of the scientific and technological background we deem at the heart of safety critical CPS and CPSoS engineering.

Digital real-time control has been key in the constant progress of embedded systems since the 80s. However, up to end of the 90s control engineering has been used on a physical domain per physical domain basis: mechanical control, electrical control, flight control etc. Multi-physics modeling and control (e.g. Modelica [Mod96]) has emerged recently, mainly necessitated by hybrid car design and enabled by Gbyte-RAM computers[1]. CPS enhance the multi-physics multi-control trend. With CPS, multi-physics control, also named generalized control, becomes the norm. Most importantly, it no longer remains *local*, as with embedded systems. It scales up to the *global*: from stamp-size SoCs up to smart cities, power grids or satellite constellations.

## 1.1 The Control-Compute-Communicate trilogy

One often defines cyber physical systems as systems where “physics, computation and communication are in interaction”. We follow this symmetry-inclined line of thought. In the sequel, we subsume “Physics” by “Control”. Control has historically developed as the science of governing physical devices (e.g. J. Watt’s flyball governor). However, with software-based virtualization of some physical and hardware resources like networks and computers, control has also pervaded the digital.

---

[1] Needed for the sophisticated computer algebra transformations of Differential Algebraic Equation systems (DAEs) into numerical integration programs.

"Feedback scheduling" of millions of virtual machines over hundreds of thousands of cores, in other words "*cybernetics*-scheduling" is now commonplace in cloud and edge computing architectures (e.g. Docker and Kubernetes[2]). Migration of safety critical command and control from dedicated servers to cloud-based platforms are underway for manufacturing, power generation, railway infrastructures, etc. (e.g. [DBC20]). In addition to infrastructure cost reduction, it aims at infinite and seamless scalability, also named "elastic machine" concept.

Controlled execution and storage platforms originally developed for IT to support B2C, B2B and social networks, are being adopted for OT, highest criticality levels inclusive (.e.g. SIL4 in railway). In addition to permeating computing infrastructures, control, and as a matter of case *decentralized distributed control,* has permeated connectivity and more precisely *virtualized* [3]networks. The sense-compute-actuate loop has been introduced to adaptively manage traffic loads. What was formerly concrete and *static,* has become virtual, dynamic and controlled. Under the influence of the ongoing IT-OT convergence, the physical and the digital are now symmetrically under control of the "governing art".

Dynamic allocation of virtualized computing and networking resources introduces new timing variability sources and availability events, which in turn may necessitate *application*-layer adaptation to these executive-layer variabilities. What was formerly stratified and addressed at design-time may become coupled and addressed at run-time through *resource-aware control*. It necessitates new applicative-executive co-engineering techniques (as an example see [Dol17], for the linear control case or [Aub10]).

Finally yet importantly, the Observe Orient Decide Act loop (OODA), that structures any supervisory level, is a control notion, a global governing structure applicable to any scale of integration: power grids, search & rescue systems, smart cities, crisis management systems etc. Control matters also for the IT & ISR[4] side of CPS and CPSoS. The techniques we put forward in section 4 (4.1 and 4.3) are intended for both the IT side and the OT side of the safe & secure continuum needed by the IT-OT convergence.

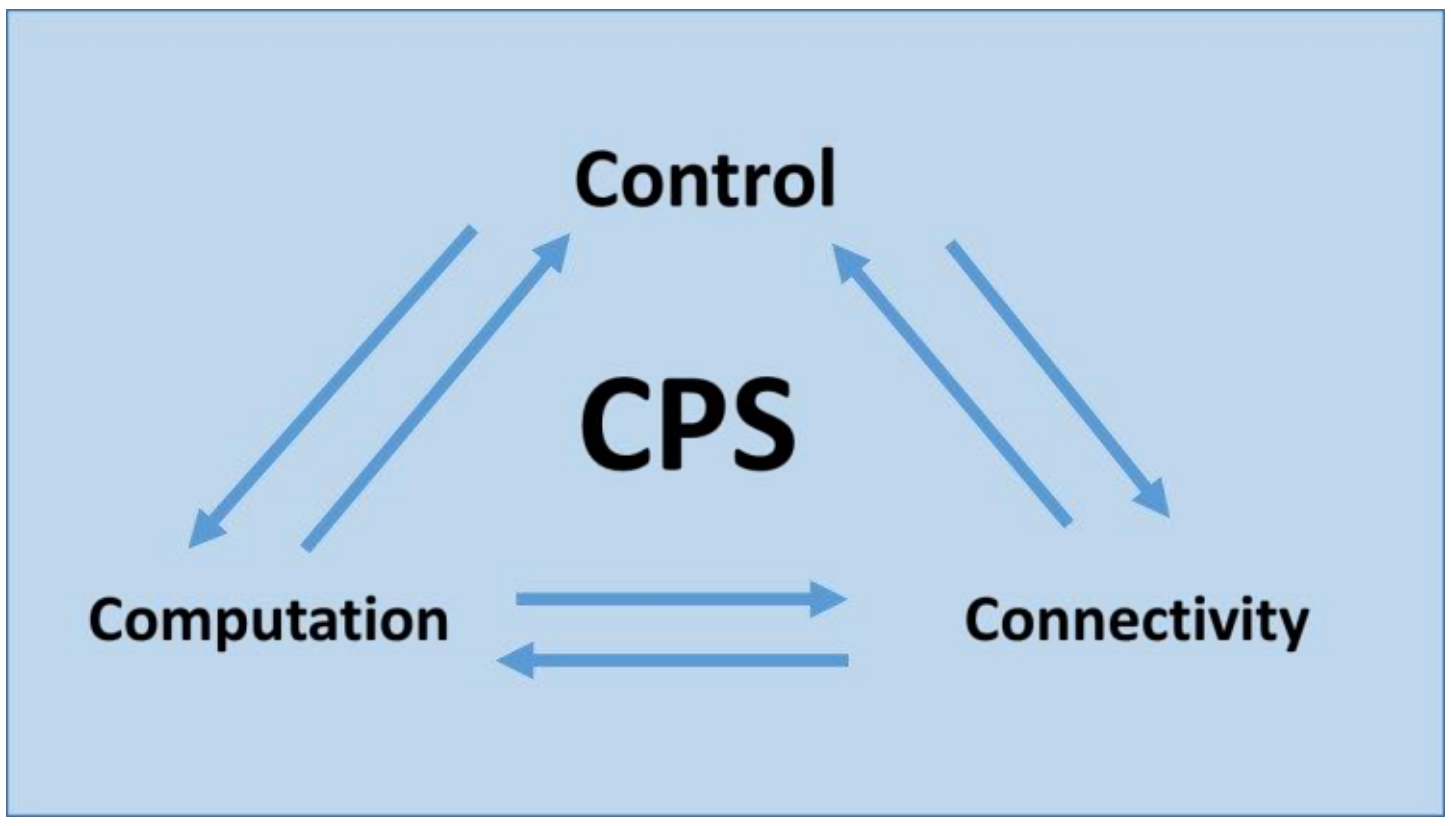

*Figure 2 – The CPS-CPSoS specificities w.r.t. to embedded systems and process automation. The governed-digital-physical continuum: symmetry of interaction between Control – Computation – Communication. Control here implicitly, like historically, encompasses the physical, and most often implies safety criticality.*

[2] Kubernêtiké again.
[3] Also named Software Defined Networks (SDN).
[4] Intelligence, Surveillance, Reconnaissance

We review the six interactions of the trilogy to underline where CPS and CPSoS depart from embedded systems.

**Computation for Control**

Sensing and signal processing have always been computation-intensive in embedded systems. Think of radars for instance. They have kept being so in spite of more than six decades of Moore law. Today the need for high performance computing (HPC) is still increasing, drawn as it is by the needs of AI-based machine vision (CNNs, DNNs[5], etc.), a "must-have" for autonomous systems. Sensor redundancy (radars, LIDARs, cameras), sensor fusion, image and video processing, NN[6] training and generalization, all have boosted the need for HPC to support the sensing part of control.

Autonomous vehicles (cars, trains, tramways, drones, underwater autonomous vehicles etc.) critically depend on visual sensing and vision-based control, both highly computation intensive. We regard autonomous vehicles as embedded systems when considered in isolation, and as CPS, CPSoS or SoCPS when connected and coordinated. The *control scope* and its life-cycle management make the difference, as proposed in 1.2.

**Control for Computation**

This notion seems to have appear with low power (e.g. mobile devices) and data-centers (orchestrated virtual machines). In both cases, control senses the available resources and the load profile, to allocate the resources and schedule the tasks in compliance with the QoS and resource footprint objectives. Operation research algorithms optimize resource allocation. Consequently, timing analysis of the executive layer, key for control stability, has become significantly harder, and safety demonstrations in turn.

**Connectivity for Computation**

Parallel and distributed computing architectures are ubiquitous in CPS for sensing, Big Data, or to support coordination of entities by means of wireless networks (LTE, 5G, …). Examples of connectivity-intensive computation infrastructures are manufacturing plants of industry 4.0 or command and control of transportation infrastructures (ATM/ATC, railway, …).

**Control and Computation for Connectivity**

Virtualization of the network resources (Software Defined Networks) has paralleled virtualization of the computing resources. It has enabled optimized sharing of the physical resources to support transparency of technological heterogeneity and to support massive transfer services with high bandwidth variability. SDN relies on *feedback scheduling*, i.e. continuous sensing of the physical and logical resources' sates to adapt the configuration and the operating modes.

**Connectivity for Control**

A crane is a CPS and an embedded system[7]. We regard a (4-crane)-crane, i.e. four coordinated cranes that lift altogether a load whose weight is beyond their respective capacity, as a CPS that no longer is an embedded system. Collaborative cranes give just an example of collaborative missions. The equivalent for defense systems, NCW[8], appeared in the 2000s. Swarms of coordinated drones is another typical example. Data-links, e.g. ADS[9]-B, CPDLC in aviation, LTE, 5G, RFID, and all types of wireless communications have enabled design of collaborative systems that go far beyond mere interoperability, interconnection, remote monitoring, or remote maintenance. It has enabled *collaborative control*, whether centralized or decentralized. Thanks to generalized wireless connectivity, new physical hazards dependent on *global* control loops are being created at all scales of integration (tight-coupling) or interoperability (loose-coupling).

[5] Convolutional Neural Networks, Deep Neural Networks

[6] Neural Networks

[7] We support the idea that the definition of CPS should extend that of embedded system, without excluding them from being (limit-case or "degenerated") CPS.

[8] Network Centric Warfare

[9] Automatic Dependent Surveillance-Broadcast, Controller-Pilot Data Link Communications

We now review some CPS aspects where engineering is challenged. They motivate the three open problems and associated research tracks proposed in this paper.

### 1.1.1 New challenges for the safe and secure

In embedded system engineering, safety is mainly addressed as *architectural mitigation* of component failures, i.e. as fault-tolerance of reliability events. For CPS engineering, we deem necessary to address safety primarily as a *controllability* issue, controllability under disturbances and uncertainties [Lev12][10]. Among the many disturbance sources, physically initiated component failures remain a major one. However, with software-intensiveness and connectivity-intensiveness, the flaws at system specification level, i.e. the risks of overlooked *interaction* failures at design time, tend to be a concern on par with that of *component* failures.

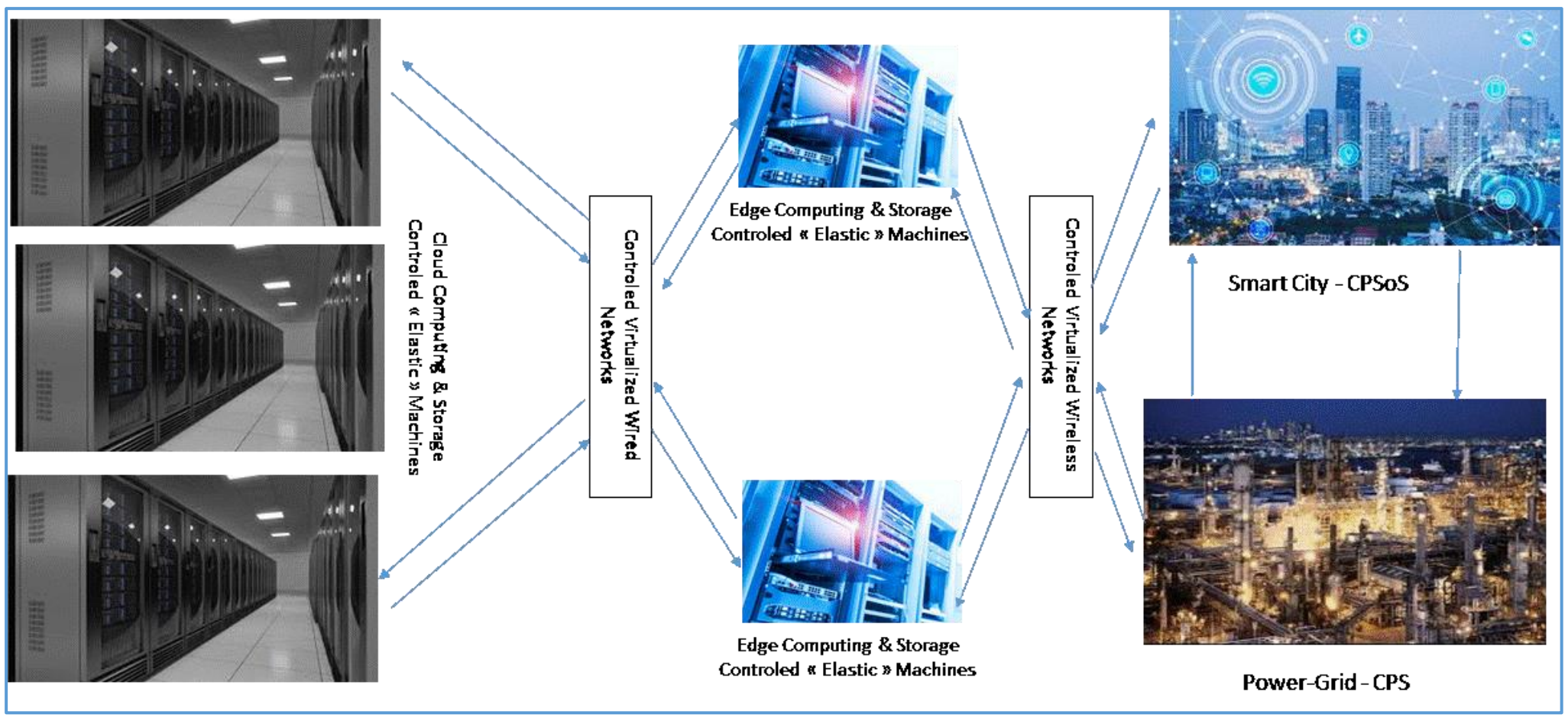

*Figure 3: pervasive control in IT-OT convergence – CPS/CPSoS: software intensive and connectivity intensive systems and systems of systems*

We observe with some concern the headlong rush towards generalized connectivity and interaction of mobile devices, machines, vehicles, infrastructures, and people. Remote monitoring of physical installations, predictive and remote maintenance are among the main business drivers of this race, amplified by the Big Data and AI-Machine Learning promises. However, even with state of the art engineering, can we trustworthily verify such CPS and CPSoS? What is *behavioral* verification *coverage* for distributed physics, systems, and systems of systems?

### 1.1.2 The logically separated over the physically shared

The promise of virtualization in OT is cost reduction, facilitation of deployment over massive numbers of end-points, and better adaptivity ("elastic resources"). However, virtualization has major drawbacks for safety assurance. First, it complicates substantiation of execution integrity. Second, sharing multiplies the single points of failures where traditional engineering cautiously resorted to *physically*

[10] N. Leveson uses the term « system-theoretic », while putting forth "control-loop structures". We deem "control-theory" a theoretical setting more beneficial for safety than "system theory". Both converge on the core idea of "state governance".

*and digitally* independent redundancies. Third, it pushes in open world what formerly was in closed world. It creates a surface of attack on the control-sensitive execution machinery.

### 1.1.3 From mono-domain centralized control to multi-domain and distributed control

It has been a long time since sensors have been shared and fused to augment observability. Collaborative observation is going a step further with CPS and CPSoS (e.g. sensor networks). More recent is joint actuation, joint control of different physical domains to the same end. Control here has to draw additional services from *fault-tolerant distributed algorithms,* for instance to resist packet drops or delays in distributed estimation problems (asynchronous reconstruction of a "virtually synchronous" global state from distributed local states).

Classical control theory, either linear or non-linear, addresses the *centralized synchronous* case. CPS and CPSoS open new fields in control: the asynchronous distributed, centralized or *decentralized* cases[11]. Cybersecurity protocols, intermingled with collaborative control protocols, worsen the complexification induced by fault-tolerant distribution.

### 1.1.4 Integration-based development .vs. specification-based development

Reuse of COTS[12] has soared in software and hardware engineering. IoT engineering is a matter of architecting the integration of existing software and hardware components in which there exists enough openness to customize their configuration and add new functionalities. Regarding mastery of safety, reuse induces limited modification capability on the components of a system. It may be a source of risk. This limited specification and implementation capability of "bottom-up" engineering is, to our knowledge, a distinctive feature that delineates the frontier between systems and systems of systems.

## 1.2 Why CPS, CPSoS, and SoCPS?

It is acknowledged that in theory the concept of System of System (SoS) should not exist, but that in practice it does when the systems that compose the SoS keep their own *goals* and *managements*. In such cases, the ability to federate the component-systems to achieve the SoS goals is limited, which may be a source of risk when safety is concerned at SoS level, i.e. when there are hazards proper to interaction at SoS level.

[Dan16] classifies the SoS by means of *control type* at inter-system level: "directed", "acknowledged", "collaborative", "virtual", and "discovered". "Directed" is the strongest control type (tight coupling) and "discovered" the weakest. We follow the idea that the type of inter-system control is a discriminating characteristic to classify SoS, and propose a characterization of the differences between CPS, CPSoS and SoCPS along this line. Such a characterization is still a matter of debate; we do not claim to put forth a conclusive argument.

First, what could be the difference between CPSoS and SoCPS? We suggest that if "CP" is placed in prefix position it means prevalence. Thus, we interpret CPSoS as CP-SoS or CPS-oS, with the intended meaning that physics is *globally* controlled by the interacting component-systems. In that case, there should exist at least one control loop that spans over two component-systems or more, whatever the [Dan16] type of control in these loops.

Conversely, So-CPS would mean that regarding *physical* control (.vs. *informational* control), all the loops are designed and properly managed within the component-systems' life-cycles. In other words, the goals at SoS level are informational, not physical. Control loops may exist at SoS level, but they actuate exclusively digital resources. From a safety perspective, there should be no issue with So-CPS since the hazards are related to physical damage[13], and the potential physical damages depend on a single component-system in spite of its communications with the other component-systems.

[11] Control by means of multi-agent systems is an example of the asynchronous decentralized case.
[12] In the generic sense of "component": a product, a software item, a hardware IP, a sub-system, etc.

Second, what could be the difference between CPS and CPS-oS? The "-oS" denotes limited capability of federating the component-systems in spite of existence physical-control loops that span over them. From safety assurance standpoint, this is potentially a dangerous situation. Contrary to the straightforward and common interpretation, we suggest that there should be no difference of *scale* between CPS and CPSoS[14]. A smart city, or a power grid may be regarded as a CPS and not as a CPSoS as long as a prime contractor, or any institution playing the same federating role, has the ability to master design and component-wise life-cycles wherever safety critical physics is governed. Global inter-system control is likely to be of the "directed", "acknowledged", or "collaborative" types since safety-criticality is at stake. When there is too much independence between the development and lifecycle management of the component-systems, there is potentially a lack of behavioral consistency on the physical aspects. One may use CPS-oS instead of CPS to manifest weaker coherency enforcement on the components, safety-sensitive limitation on federating management.

## 1.3 Synopsis

We have proposed control, primarily of the physical but also of the digital, as a key characteristic of CPS, and *physical control scope* as an indicator to discriminate between CPS-oS and So-CPS. Whatever the scale of integration or interoperability, when physics is in interaction through the component-systems, we propose to choose between CPS and CPS-oS according to the answer to the following question: at inter-system level, is control under control?

What is at stake with the emerging digital society has been summarized in figure 4, that we have borrowed from [TCI21] and augmented slightly. We have supplemented the digital continuum with the *governed digital-physical continuum*, to manifest pervasive IT-OT control with all its shades of business, safety, and security criticality. We regard the informational-physical control continuum as a candidate central issue, if not concern, of the digital society underway.

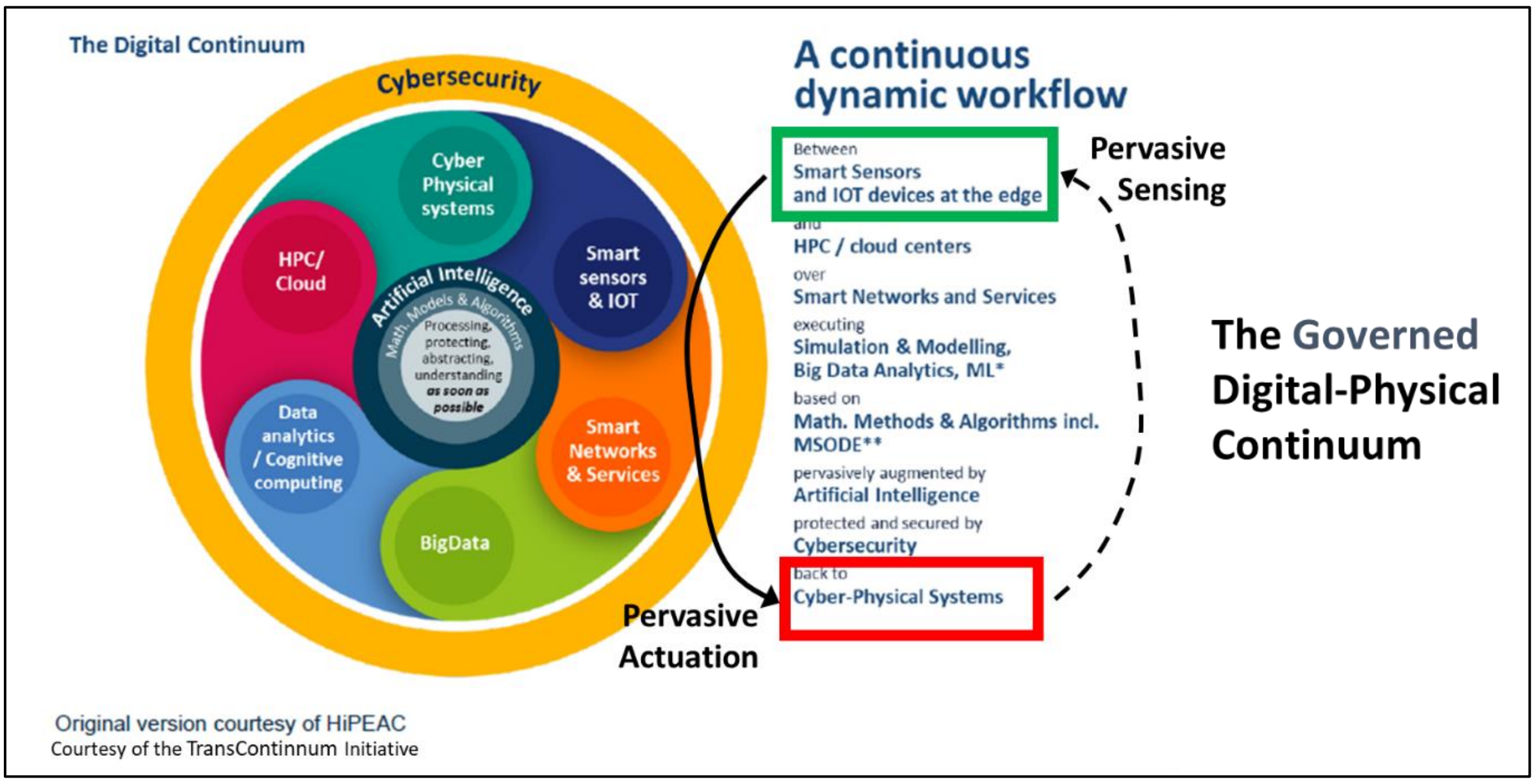

[13] The case would be different if safety were extended to business damage
[14] The concept of system is recursive and scale-invariant

*Figure 4: A conceptual reformulation of figure 3, as an augmented version of the IT-centric view of [TCI20]. The continuum is Digital-**Physical**, IT-**OT**, the two components interacting through controlled or uncontrolled digital-physical influence networks. We have used the STPA graphical notation for control loops. Besides AI-Machine Learning, progress in distributed fault-tolerant and resilient control is needed to build a trustworthy digital society. Any accident is primarily a loss … of control.*

# 2. Three Open Problems in CPS Engineering

We focus on three engineering problems we think are important to address the new IVVQ and safety engineering challenges that arise with CPS and CPSoS. They are all related to *behavioral analysis*. The raise of rework and "emergent properties" correlated to that of distribution and networked control motivates the first one. The lack of measures of behavioral verification coverage at system level is the second one. The third open problem is the widening fidelity gap between safety assessment models and system engineering models.

AI-enabled autonomy in passenger transport is a game changer for CPS and CPSoS IVVQ. In case of accident and legal consequences, the balance of responsibility options between the faulty-machine and the faulty-operation will no longer be the same [Frz12]. We perceive some recent renewed interest for formal verification techniques on development of autonomous systems. A similar phenomenon occurred two decades ago with the advent of new cybersecurity threats in IT and open-world OT. In spite of the new scales of complexity addressed in section 1, leveraging formal specification, design and verification techniques at system level, primarily on safety critical control aspects, is the rationale that underpins the next three sections.

## 2.1 Compositional verification & certification (Pb#1)

In software engineering, modularity has been instrumental for scalability. Do we have any equivalent for CPS and CPSoS development, verification, and certification?

On the system engineering side, the notion of functional chain is the standard approach to specification and unit system testing. On the safety assessment side, dysfunctional analysis relies on two standard methods:

1. From the local to the global: identification of the component failures[15], and analysis of failure propagation i.e. of the "cascading effects" or "domino effects" (FMECA[16]),
2. From the global to the local: Boolean modeling of the causes (FTA[17]) of the macroscopic feared events identified by risk analysis.

Both analyses are chain, tree, or DAG[18] based, i.e. *cycle-free*. As explained in the previous section, CPS and CPSoS highly depend on control. They are *cycle-intensive.* Control loops are causality cycles not causality chains, trees or DAGs.

Therefore, control is difficult to modularize because of this twofold cyclic nature:

- *Spatial*: the data-flow dependencies between the control loop entities form a spatial cycle (sensors -> controllers -> actuators -> plants -> sensors),
- *Temporal*: these spatial loops compute state updates that are iterated over time, thus constituting temporal loops.

---

[15] Random and systematic. For the latter (i.e. residual development faults, in other words development assurance failures), the initiators are prototypical representatives of unknown potential errors.

[16] Failure Mode Effect and Criticality Analysis

[17] Fault Tree Analyis : *all* the causes of a macroscopic feared event are asked by the safety norms to be represented by a Boolean combination of the anticipated possible component failures at meso- or micro-level.

[18] Directed Acyclic Graph

This space-time twofold cyclicity tightly couples the systems' functions and components over possibly deep spatial and temporal horizons. When one modifies some system, how to substantiate with strong arguments that one masters the space-time propagation effects without performing global regression testing campaigns? As one needs IVVQ facilities as close to the final implementation and operational context as possible for these campaigns to be valid, their cost are prohibitive. Hence the importance of synthetic environments and digital twin models. However, they are only partial answers to verification coverage of infinite and highly complex behavioral spaces.

The situation worsens with distribution. The size of the global reachable state space to explore grows exponentially with the number of networked components, and with the size of their respective reachable state space[19]. The more distributed, the more *illusory* sufficient verification coverage by model simulation and on-target testing.

We need disruptive methods of *compositional* IVVQ and *verification-by-design* to overcome the new scales of behavioral complexity.

## 2.2 System verification coverage measures (Pb#2)

Rework has tremendous economic impact and may result from many reasons. Insufficient behavioral analysis at specification, design and V&V time is a major one. How to decide *verification coverage sufficiency* at system level?

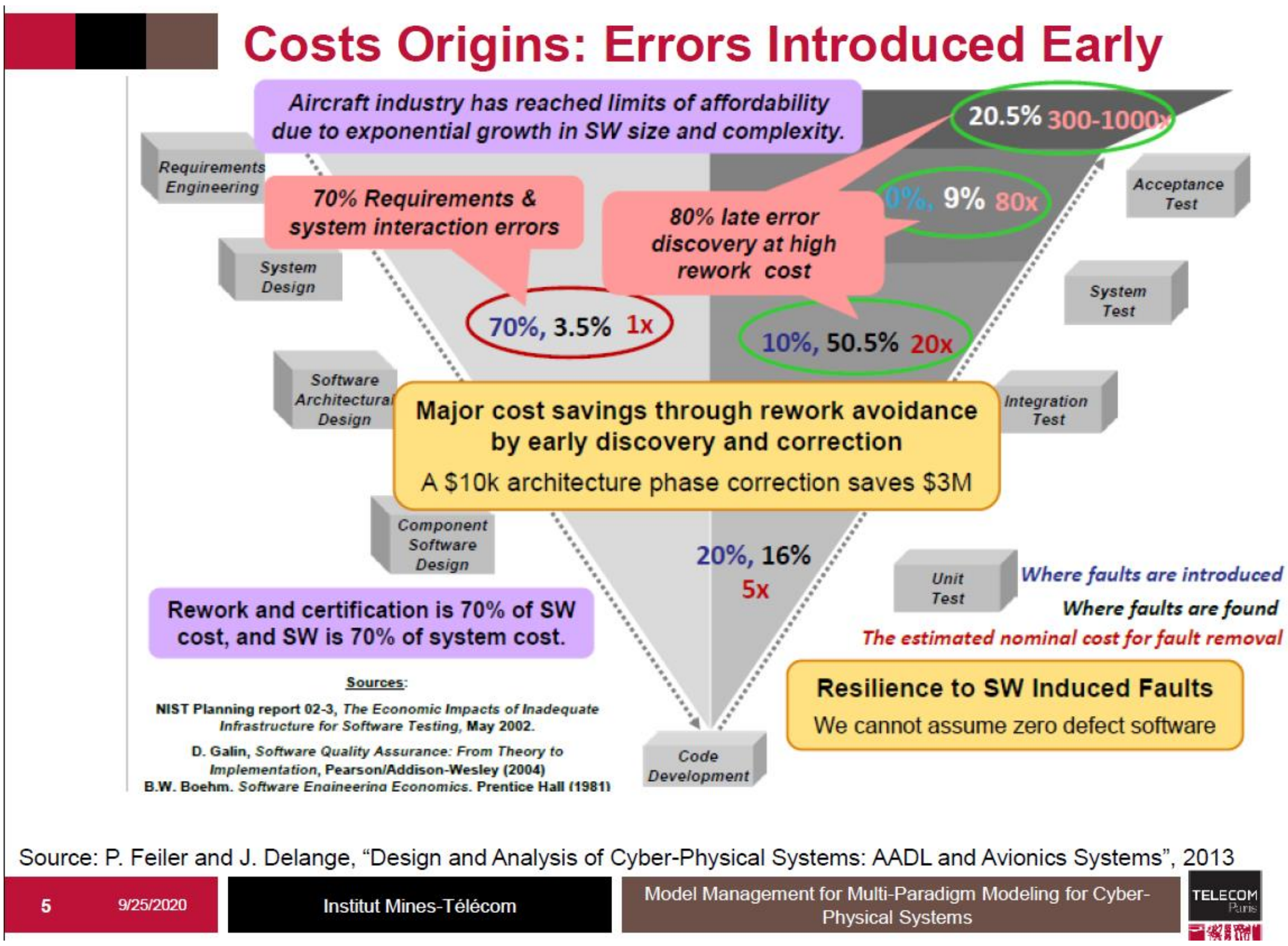

*Figure 5 – We reuse this figure, while reversing the claim: 70% of system cost stems from loose **system** specification, which in turn leads to significant rework at software specification level. On software-intensive parts of systems, system and software specification may locally coincide.*

For the digital, i.e. hardware and software, there are two coverage measures, applied in order:

[19] Size is cardinality for finite-state systems, and undefined for infinite-state systems (e.g. physical systems, software). Exponential growth surges from parallel products (input vectors, parallel composition) and worsens continuously from synchronous products to asynchronous products.

1. *Requirement coverage*: the operational, functional and organic requirements are numbered and stored in requirement databases. The verification test plan defines as many test cases as needed to exercise each of the requirements sufficiently. 100% coverage of the requirements is commonly required for the higher levels of criticality. This coverage measure is enforced by means of traceability links between the requirements and the test cases,
2. *Structural coverage*: we anchor activation counters to the elements of the structure to record how many times the previous V&V activities have activated them. The aim is to detect the parts left inactivated after[20] completion of the *requirement-based* testing campaign. These potentially not "truly dead" parts are regarded as sources of emergent properties, named unintended functions in aeronautic safety standards.

Structural coverage analysis is a *behavioral* coverage analysis *substitute*. It is one of the most important rigor modulation means of the safety assurance standards for all industrial domains (e.g. the DAL-dependent IC, DC, MC/DC coverage criteria). It is one of the most fundamental means of building trustworthiness for the digital. Even though nature solves variational invariants, nature is not a computer [Cop17]. At least we do not have access to instructions on which we could hook activation counters, we have no behavioral coverage substitutes. How to provide behavioral coverage measures for the physical, and for the governed physical?

## 2.3 Safety assessment on high fidelity models (Pb#3)

There is no need to add the complications of CPS w.r.t. embedded systems to encounter the third problem. It was identified for embedded systems by the early 2000s [Ake06]. Its first aspect consists in the split between functional safety engineering and safety assessment at modeling & analysis level. System Engineering (SE) uses behavioral modeling and simulation. Safety Assessment (SA) uses FMECA and FTA models[21]. For highly integrated software-intensive and distributed CPS we have questioned the validity of FTA models [Led20]. The fidelity of MBSA models w.r.t SE models, though better than FTA models', is still a matter of concern[22]. The situation is worsening with CPSoS. The "reality gap" is widening.

The second part of the model fidelity problem is a matter of risk analysis mindset. We follow [Lev11] and support the system-theoretic approach to risk analysis. We privilege the term "control-theoretic" to that of "system-theoretic" because we intend to draw more from control theory than from general system theory (c.f. 4.1 on control invariants and 4.2 on observability). The basic view is the same: analyze accident risks primarily as *control losses*. Function/component losses or malfunctioning are *causal factors* of control losses.

[20] As opposed to *during.* There is no structural coverage goal. The goal is *detection* of the possible useless implementation parts. They are deemed to constitute risks of harmful "emergent properties". The goal of structural coverage analysis is *not* structure activation by any means.

[21] On average in industry, as of writing this paper. Safety assessment may also resort to Model-based Safety Assessment (MBSA), to Monte Carlo Markov Chain estimations, etc. [Bou08].

[22] For instance, synchronous abstraction generates causality cycle errors in SA models because of the twofold cyclic nature of control.

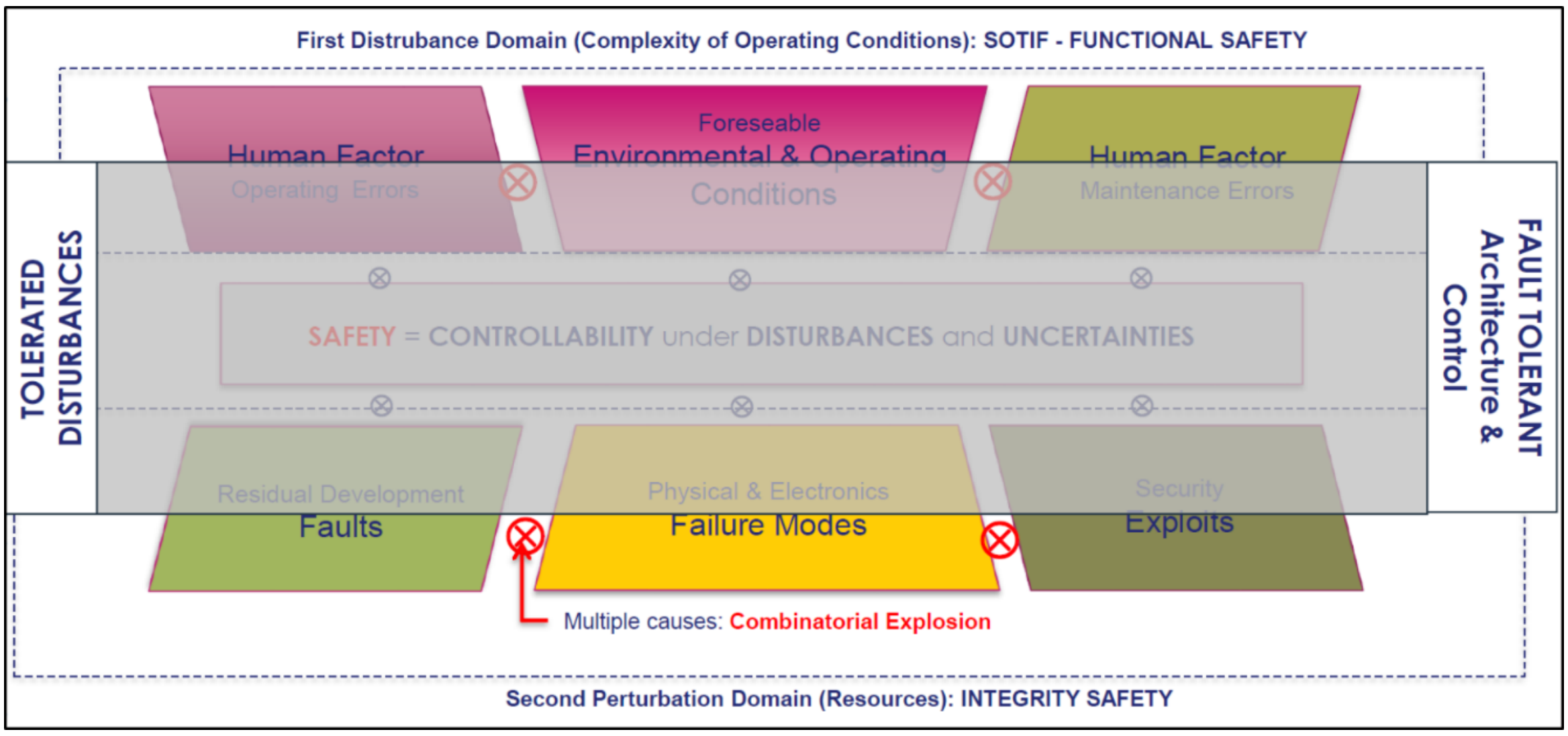

*Figure 6* – System safety as *control game.* Three types of internal adversaries (integrity safety) are represented on the lower half. The component failures and their associated "domino effects" analyzed by FMECAs constitute the prominent contributor. Among the external "adversaries", many (unintended or intended) human errors, and many unfavorable environmental conditions (SOTIF). The six sources interact with the nominal behavior (first 6-fold parallel product), and superpose with one another (second type of parallel product). The grey area notionally represents the part of disturbed behaviors that remains controllable by the FDIR[23] and control players (controllability domain). Only low-order disturbance superpositions can stay in the controllability domain.

Component failures and the ensuing failure propagations remain a major source of potential control losses. For instance on a medium-size aircraft one counts about 6500 component failure modes handled by crew and/or maintenance. However, inter-system specification flaws and human factor design errors are growing concerns. They lead to *interaction* failures, possibly without any component failure in the causal dependencies. Hence the crucial need of applying control-oriented risk analysis methods like STPA[24] on CPS and CPSoS. It starts with control of safety-constrained states, not with local breakdowns, or with dubious causality-inverse modeling of accidents.

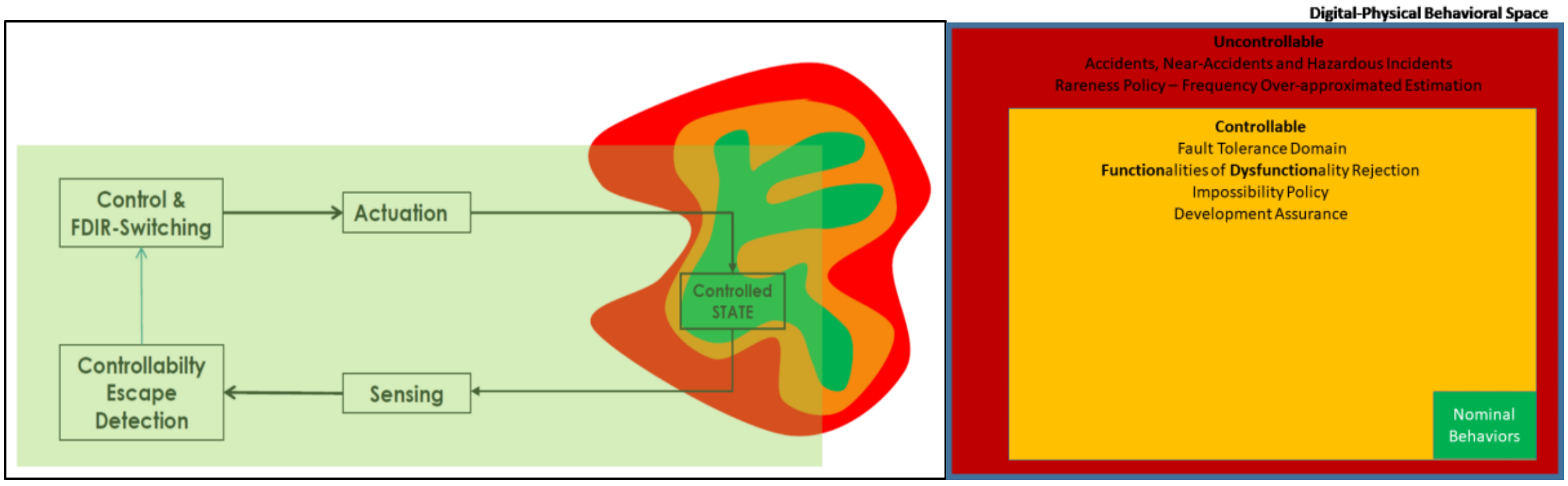

*Figure 7:* the control game consists in keeping the state subject to the safety constraints within the green zone (the nominal), or at worst within the amber zone, i.e. the robustness zone where dynamics should stay thanks to the degraded modes and the FDIR mechanisms. Out of the FDIR controllability domain, the system enters "red zones" of the accessible state space, i.e. accidents of any kind of severity. Red means contract violation, not catastrophic effect.

[23] Fault Detection Isolation Recovery
[24] System Theoretic Process Analysis

Safety assessment relies on FTA for DAL assignment, for single cause failure detection (order 1 minimal cutsets or sequences), and for feared event probability calculation. Originally conceived for *structural*[25] dysfunctional analysis, FTA has been progressively extended, in usage, to *behavioral* dysfunctional analysis. The characteristics of this extension are the following:

1. *Inverse mode*: back from effects to causes, down from global to local, on the digital and on the physical, and on the twofold cyclic governed digital-physical,
2. *Set-based:* the root event denotes infinitely many red-zone behaviors (initial conditions of the backward analysis),
3. *Abstract interpretation:* software behavior and physical behavior are addressed qualitatively like in qualitative physics [Kle84],
4. *Dysfunctional:* it decomposes global dysfunctional behaviors into interactions of more local dysfunctional behaviors,
5. *Boolean-definability:* the interactions between the dysfunctional behavior-components are all representable in Boolean algebra, i.e. without time, sequence, values or cyclic dependencies,
6. ***Completeness****: all* the combinations of initiators that may lead to occurrence of the top level feared event are claimed to be represented in the Boolean tree. In case of incompleteness, the root probability is ***underestimated***.

When used beyond the structural on CPS and CPSoS, this analysis is so “powerful” that its validity has been collectively questioned [Led20]. We need an alternative approach. How to work in *direct* mode, with behavioral modeling languages richer than Boolean combinational logic? How to take into account the system engineering *high fidelity* behavioral models in safety assessment? For the finite-state case [Ake06] has explored some paths in that direction, fully and manually abstracting from the governed physical. How to limit manual creation of SA models to the situations where no system engineering behavioral models are available?

On high-fidelity behavioral models developed by system engineering, we would need for safety assessment probabilistic quantification of:

1. safety, mission reliability, and availability events,
2. FDIR detection failures,
3. FDIR recovery failures,
4. Functional failures of IA-ML-based functions,
5. And more generally of functional failures dependent on randomized algorithm or statistical estimations.

As explained in 4.3, some solutions exist and are in practice [Gob13], [Mor16]. Notwithstanding the accelerated Monte Carlo methods, the scalability-fidelity trade-off for the systems addressed in section 1 remain an issue [Bou08]. We propose in sections 3 and 4 to explore compositional predicate abstraction of hybrid system models to go further [Mov13], [Slo13], [Bog14], [Bou15], [Sao18a], [Sao18b].

# 3. Bridging the Gap: the µΩ Open Innovation Project

## 3.1 Why µΩ?

Does academic and industrial state of the art now enable us to engineer a small, yet representative, CPS/CPSoS exactly as we wished to? If one mobilizes the above mentioned theoretical background, the available modeling languages and verification tools, can we get full mastery of the governed digital-physical behaviors, at affordable price?

[25] Dependence network analysis, with loss and malfunctioning semantics of dependencies

Such an ideal CPS/CPSoS engineering we name it "Ω-engineering". In spite of significant progress in hybrid system modelling, simulation, and formal verification, in spite of mature software and hardware development tools, we tend to consider that CPS/CPSoS engineering is still closer to α-engineering than to Ω-engineering.

"µ" refers to "micro". The µΩ project aims at reaching Ω-grade engineering on a µ-CPS/CPSoS, to start with, *temporarily*[26] getting rid of the scalability constraints. The goal is to explore the most disruptive techniques that could potentially meet the three aforementioned challenges.

## 3.2 The origins

### 3.2.1 The Overarching Properties Work Group (OPWG)

A joint EU-US group of avionics certification experts, academic researchers, software vendors and certification Authorities, has worked from 2015 to 2018 to design a new approach to system, software and hardware certification. To test the viability of this new assurance method, an open source use case named µXAV based on a "µ-aircraft" that embedded three coupled systems has been used [Led17]. The µΩ project reuses the µXAV use case, and extends it. Its primary concern is now CPS engineering for any scale of system (from SoC to CPSoS), while keeping emphasis on safety engineering and cost-effective certification.

### 3.2.2 The Embedded France work group on safety norms

Embedded France is a non-profit organization that represents the French CPS ecosystem w.r.t. to the French ministry of industry [EF20]. It coordinates a few work groups. One of them is devoted to the cross-domain comparison of system and software safety standards. When needed, it uses the µΩ use case to concretize issues debated on safety assurance principles.

### 3.2.3 The "Chaire ingénierie des systèmes complexes"

Founded by Ecole Polytechnique and THALES by 2003, this corporate sponsorship program aims at supporting research and education in CPS engineering. The µΩ open-innovation use case is meant to facilitate scientific collaboration between academia and industry on CPS engineering [ISC20].

## 3.3 The objectives

The µΩ project is a collaborative attempt to identify state-of-the-art CPS engineering at minimum effort, without scarifying scalability analysis. The main focus is mastery of digital-physical behavioral complexity at affordable price for distributed, software intensive, safety critical and security critical CPS/CPSoS. We describe µXAV, the tiny CPS/CPSoS specified from 2015 to 2018. We give a status on its current development and present its planned evolutions.

## 3.4 Overview

µXAV is a pseudo-drone whose embedded system architectures are intended to comply with large aeroplane certification regulation (CS 25 [EAS20]). These architectures must be devoid of single point catastrophic failures. To meet this regulatory safety objective, duplex architectures, health monitoring and fault-tolerance mechanisms have been introduced. It is of course not the case on true drones for obvious reasons of weight, cost, and overkill w.r.t. the true applicable safety regulation.

[26] The use case is « small » but end-to-end. Scalability will be addressed everywhere on two aspects: algorithmic complexity and staff learning curve.

AV stands for Air Vehicle, X for any kind of mission/purpose, and µ for micro as in µΩ. µXAV is a cargo drone that transport payloads. It embeds a multi-system that mimics *generalized control* (electrical, mechanical, and hydraulic). It is not representative of the kind of CPSoS reviewed in the first section. Addition of some features typical of SoS and *structure varying* systems is planned.

The repository of the 2015-2018 collaborative project is available at https://github.com/AdaCore/RESSAC_Use_Case.
It contains about 20 specification documents (operational, functional, architectural, safety), at two levels of refinement (layer0 and layer1), and a few (incomplete) models.

## 3.5 The operational viewpoint

µXAV is capable of autonomous and remotely controlled missions. Its systems ensure flight for a large domain of payloads, mission profiles and weather conditions. Range, max take-off-weight (MTOW), availability, mission reliability, and energy efficiency are the key performance indicators and values for the customer. Depending on the payload, on the environment conditions, and on load of the batteries, missions last from a few minutes to about 20 minutes.

µXAV is a *product-line*. It supports variability domains on payloads, mission vignettes, internal architectures and operating procedures.

The missions are composed of 7-phase sequences: pre-flight (mission preparation), take-off, climb, cruise, descent, landing, post-flight. Presently, the drone's dynamics is simplified. It is not a true hexa-copter; flight mechanics is restricted to the 2D-(Distance, Altitude) vertical plane and pitch. The drone is a 3D mechanical body for 0D platform design (set-based dimensioning and trade-off analysis). But it flies and navigates in the 2D vertical plane.

Mission preparation consists in defining the distance to travel, the cruise speed and the cruise altitude, the navigation mode and the navigation option. The navigation mode (A or RP[27]) defines whether the drone is autonomous, or guided by the ground station. The navigation option specifies the reference navigation parameter used by cruise regulation (speed, altitude, or energy minimization[28]). The continuous-time part of the dynamics is non-linear and switched, either by ground navigation commands, or by onboard fault tolerance reconfigurations.

There are no intermediate waypoints, and no air separation aspects, yet. Simple air separation constraints will be introduced with the SoS extensions (see §3.14). When too many failures or too disturbing environmental conditions occur, the ground station operator or the drone safety monitors can trigger emergency landing.

Emergency landing plays the role of the High Assurance Controller (HAC) defined in the simplex design pattern [Wan13]. It aborts the mission and performs a vertical landing that should preserve the payload and the drone's integrity. The mission management, energy management and flight control functions jointly play the role of the High Performance Controller (HPC). Mission Management depends on a battery charge predictor and on a range predictor developed by Machine Learning. µXAV supports µ-experiments on certification of trustable embedded AI.

For more details, see the functional specification documents on [RES18].

[27] Autonomous (A), Remotely Piloted (RP). Remotely Guided would be more appropriate as remote short-term control of the drone, i.e. piloting, is impossible.
[28] Named low-power in IoT.

## 3.6 The physical viewpoint

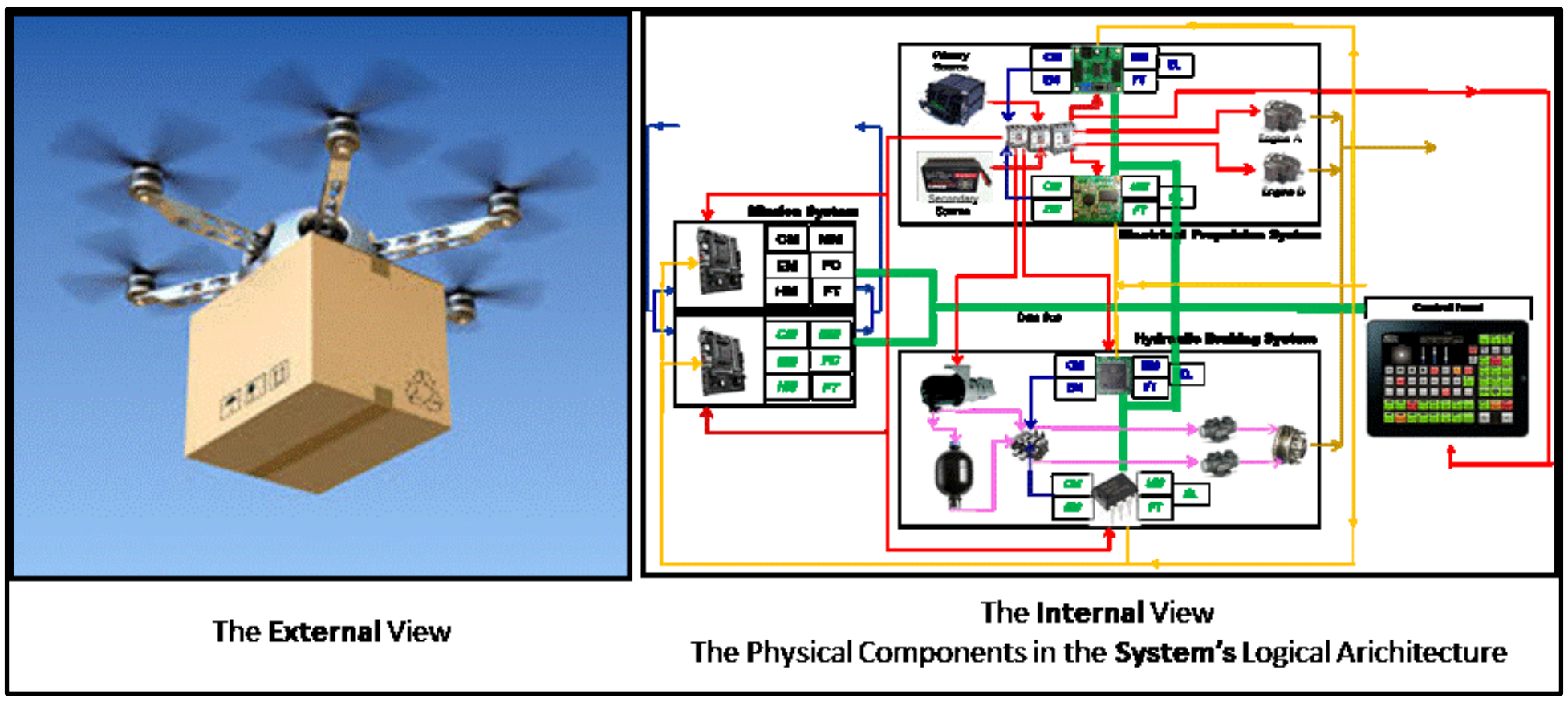

*Figure 8 –Physics of the drone (platform) and of its five systems (mechanical, electrical, hydraulic, mission management, connectivity management)*

Physical engineering at platform level comprises:

1. 3D (steady) or 4D (unsteady) modelling and simulation, to solve the partial differential equations (PDEs) of structural mechanics, fluid mechanics, electromagnetism, thermal transfers, etc[29],
2. 0D (steady) or 1D (unsteady) modelling and simulation, to solve the ordinary differential equations (ODEs) or the differential-algebraic equations (DAEs) of electrical circuits, hydraulic circuits, multiple body dynamics etc[30].

The 3D-4D engineering defines and optimizes the mechanical shapes of the drone to meet a set of objectives that condition mission performance. For instance, it computes the temperature field within the equipment and payload bays, conditioned by the cruise speed, the cruise altitude and the temperature of the atmosphere. This dynamic temperature field in turn partially conditions the battery life and the drone's autonomy: from ~20 mn full charge hot conditions to ~10 mn full charge cold conditions, everything else being equal. It also affects the electronic components' reliability and the electrical motors' durability.

Given the platform design parameters (size, weight, bay capacities,…) supplied by 3D-4D engineering, 0D-1D engineering selects the system components and predicts mission performances. The other way around[31], given a target performance domain, it computes the needed component and platform characteristics to meet the mission objectives. 0D-1D engineering is performed iteratively in two ways: first to identify the performance targets and to make a start on the component characteristics. Second, after detailed platform and system design, to refine the performance estimates.

[29] Physics of fields and waves

[30] Reticulated physics, i.e. multi-domain energy circuits.

[31] Inverse problem.

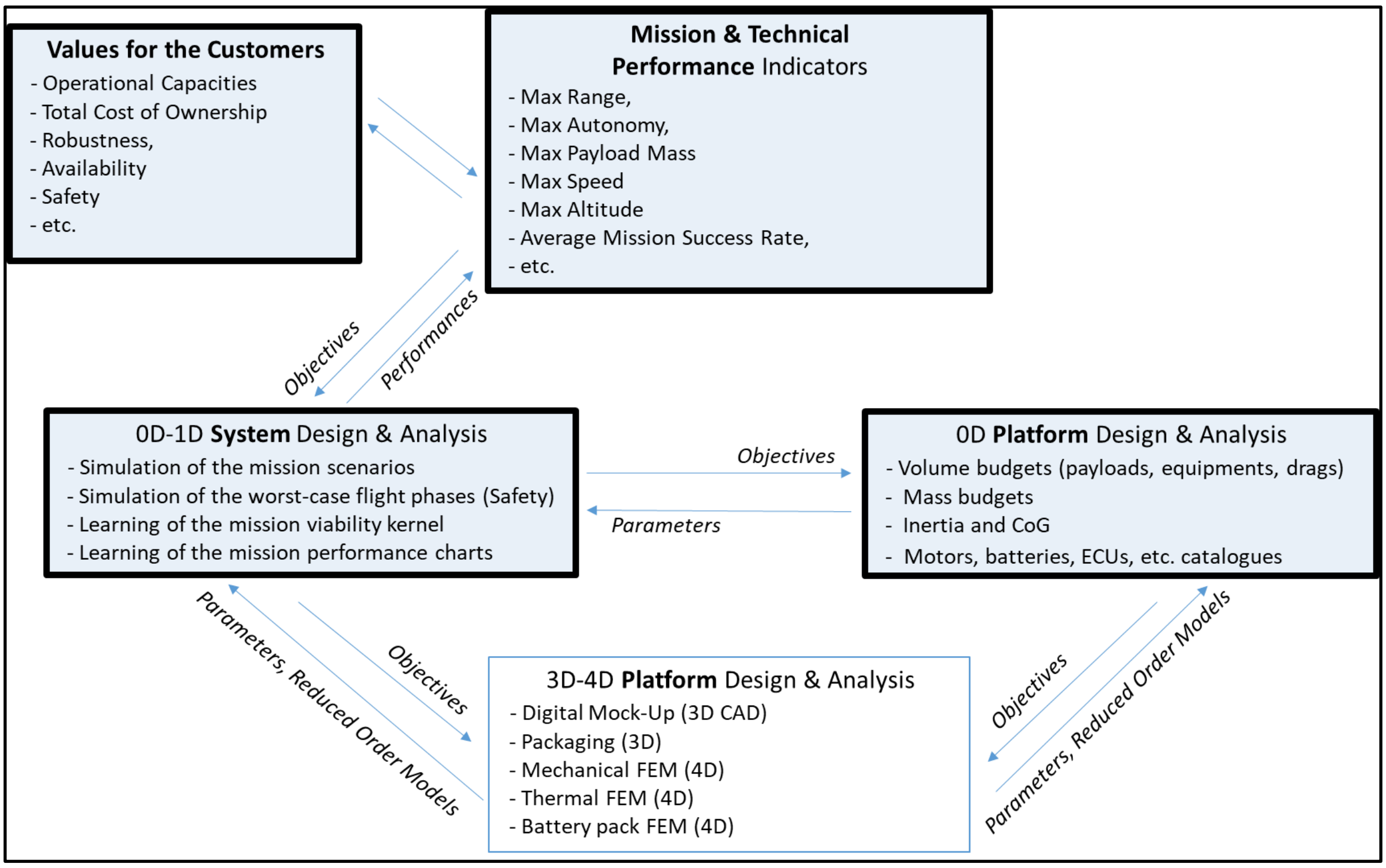

*Figure 9 - Overview of the µΩ physical engineering process. It interfaces with the platform Finite Element Method (FEM) calculations but does not address this part. Two rounds of 0D-1D analysis: 1) defining the objectives, 2) checking they are met, for a product-line configuration domain.*

The µΩ project addresses the 0D-1D engineering process and its way to interface with the 3D-4D one by means of Reduced Order Models (ROMs), surrogate models, ANNs[32], or empirical formulas. In table 1, we give the platform design parameters, whether they participate to the product-line variability domains, and their influence on the performance indicators.

| **Sizing Parameters** | **Product-line Variability** | **Influences** | **Platform Performance Indicators** | **System Performance Indicators** |
|---|---|---|---|---|
| height, width, depth | yes | drag, mass, bay volumes | cross-section, Cx usable volumes | maximal range (conditioned by payload and weather condition), take-off and landing safety envelopes |
| material density | yes | empty mass | calibrated weights | maximal range, flight level, speed limit (conditioned by payload and weather condition) |
| equipment-bay volume | yes | 3D-layout | pass/fail on a set of equipment configurations | pass/fail on a set of mission objectives |
| payload-bay volume | yes | 3D-layout | pass/fail on a set of payload configurations | pass/fail on a set of mission objectives |
| cooling air inlets | no | inner temperature field | cooling flow chart (conditioned by operating point and weather condition) | temperature-sensitivity chart on a predefined set of missions |

*Table 1 – Platform modification parameters and associated effects*

[32] Artificial Neural Networks.

## 3.7 The environmental viewpoint

The system inputs are:

- the ground station user-commands
- the field operator commands,
- the external disturbances: wind gusts and icing,
- the 35 random failure events (motors, batteries, micro-controllers, power switches, electrical pump, etc.).

See [RES18] for detail on the field operator and ground station commands.

Part of these inputs are subject to uncertainties. Set-based methods (intervals) and *worst-case* analysis address uncertainties for deterministic safety. For mission reliability, uncertainties are handled by means of probabilistic *average-case* analysis. For probabilistic aspects of safety, probabilistic modeling of uncertainties is under worst-case and extreme value regime.

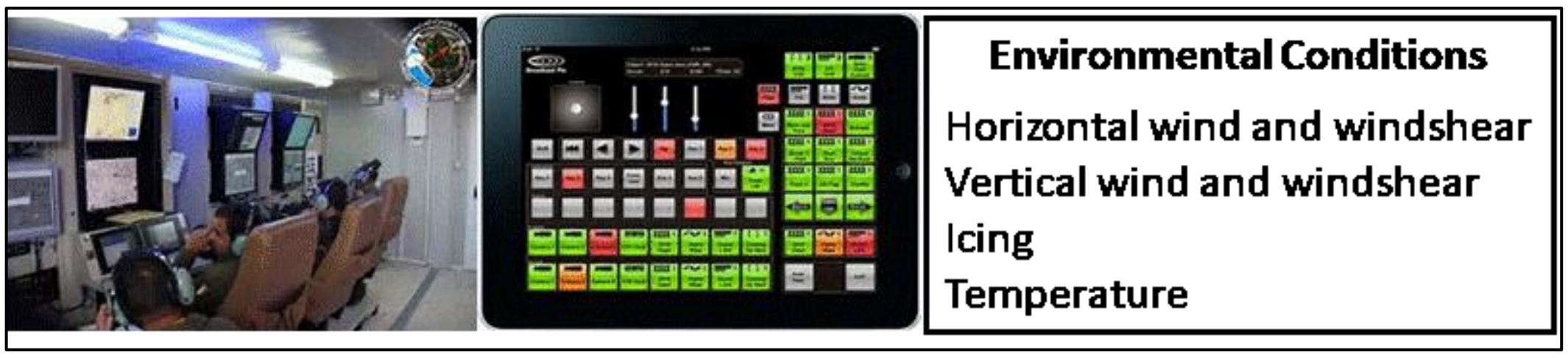

*Figure 10 –The external input sources.*

## 3.8 The embedded systems

The use case has been designed to explore engineering and assurance of a safety critical generalized control loop that span over five coupled systems:

- Mechanical platform and inertial sensors (MS),
- Electrical Propulsion System (EPS),
- Hydraulic Braking System (HBS),
- Mission Management System (MMS),
- Connectivity Management System (CMS).

The ground station was initially out of scope, but it will be integrated in the scope with SoS extensions.

The main control loop is subject to perturbations. They are the “adversaries” of the safety differential game addressed in section 2.3 and 4.3:

- **Environmental conditions***:* wind gusts, icing,
- **Human errors***:* wrong mission parameters at pre-flight, wrong navigation commands during flight; 8 different commands can be sent through the two channels (panel, station),
- **Failure modes**: 35 component breakdowns may physically initiate cascading failures. They are the *anticipated* failure modes, diagnosable by the health monitoring function (F_HM), and addressed by maintenance. There are many more possible failure modes, some of which are unknown unknowns.
- **Systematic failures**: some specific scenarios may activate unknown residual development errors (specification and/or implementation). This issue is addressed by minute design of the FDIR detectors (see §4.2), and control-oriented functional FMEA supported by STPA.
- **Fault-tolerance**: 30 FDIR[33] recovery transitions, potentially generating control instability (e.g. asynchronous delays between micro-controllers, value jumps of software variables possibly leading to transient actuation shocks),

[33] Fault Detection Isolation and Recovery

The global combinatorial complexity generated by the 35 anticipated failure modes potentially triggering 30 recovery configurations at any time during the 7 mission phases, each one possibly interleaved with 8 commands, is upper-bounded by 35x30x7x8= 58800 dynamic contexts[34]. The true number is significantly lower. The continuous wind and icing perturbations may superimpose *at any time* with these tens of thousands of dynamic contexts.

The 35 failure modes are the µ-counterpart of the 6500 failure modes mentioned §2.3. The at most 58800 cases "multiplied" by the infinitely many continuous ones illustrate the *multiplicative* combinatorial structure represented figure 6. Ω-engineering should manage to perform extensive deterministic and probabilistic analysis of this combinatorial structure, with coverage measures, performed either on the high fidelity system engineering models, or on *provably* conservative abstractions thereof.

The CS25-inspired µΩ V&V *coverage* objectives are:

- full integrity at power-on, two failures during mission, any controllable continuous perturbation at any time,
- one undetected latent failure at power-on, two failures during mission, moderate continuous perturbations at man-chosen critical times,

Figure 11 presents the unmapped functional structure and figure 14 its mapped counterpart.

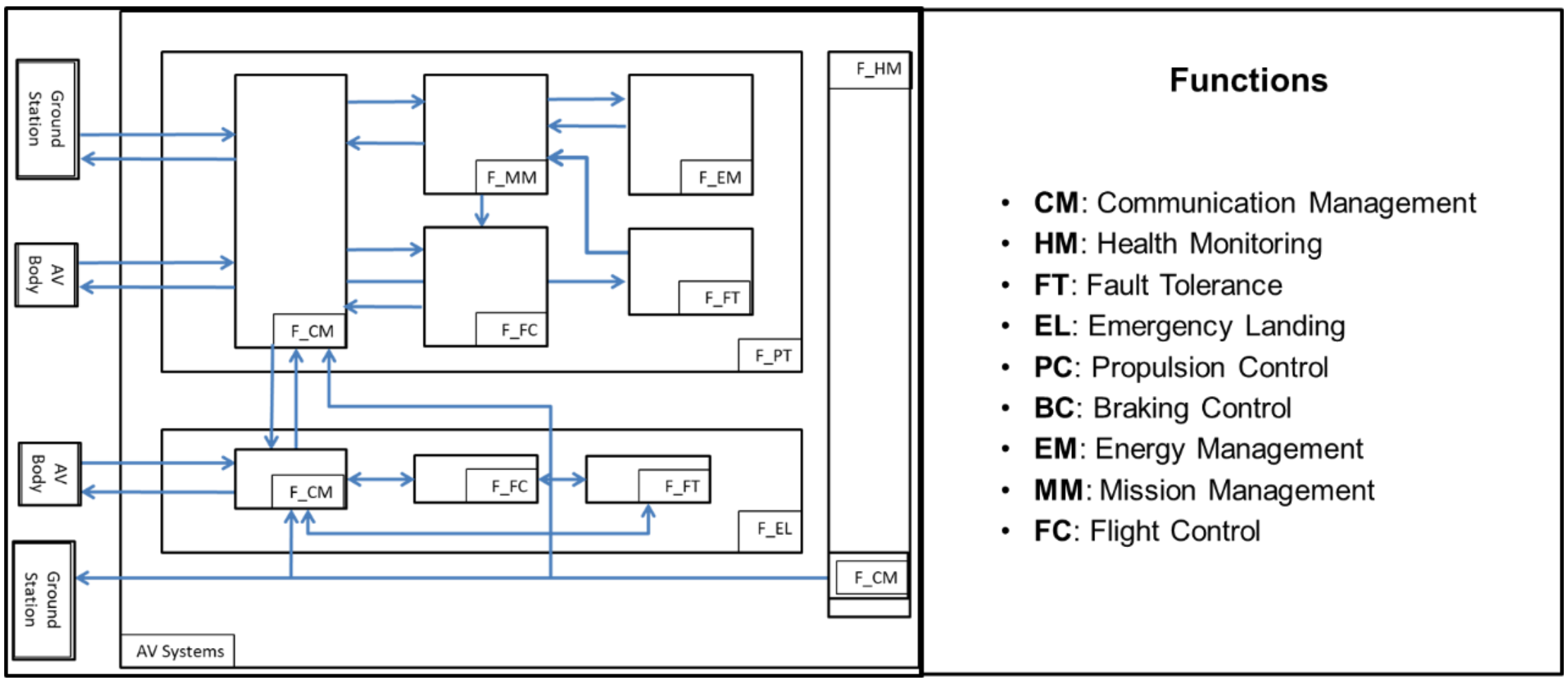

*Figure 11 – Functional architecture of the µXAV. There are many cycles in the interaction structure. The control loops are analyzed with STPA (risk analysis, functional safety specification, design constraints).*

The top-level functions are:

- Payload Transport (F_PT): it is in charge of delivering a given payload at a given distance, after cruising at a specified altitude, or speed, or in optimal energy saving mode. F_PT decomposes into five functions. Flight Control (F_FC) ensures short-term control of the drone. Energy Management (F_EM) and Mission Management (F_MM) monitor mission viability. In case of energy shortage or exceedance in environmental perturbations, F_MM aborts the mission and transfers control to F_EL. This function is in charge of grounding the drone safely. One calls "emergency landing" such a phase.
- Emergency Landing (F_EL): The drone descends and lands almost vertically, wherever it was cruising or hovering. F_EL is the most safety critical and resilient function: it can be executed by the 2 micro-processors of the MMS, and by the 4 micro-controllers of EPS and HBS.

[34] Assuming a *single* failure mode occurs during the mission

- Health Monitoring (F_HM): the electronic devices are self-diagnosed by built-in tests. The physical devices are monitored and diagnosed indirectly by “virtual sensing”, i.e. by augmented observability. Two kinds of methods are used and compared on-line:
    - **Model-based**: the specifying Simulink models used for design and V&V of the control laws are enriched with models of faults for the motors, batteries, pump, and brakes. Then, using analytic redundancy methods one can detect and partially isolate at run-time the modeled-fault occurrences [Fri17],
    - **Data-based**: resorting to the wide spectrum of machine learning techniques, deviation form normality can be detected. For frequent failures whose signatures can be learnt, one can provide some isolation and diagnostic capabilities by adding classifiers of fault-effects.

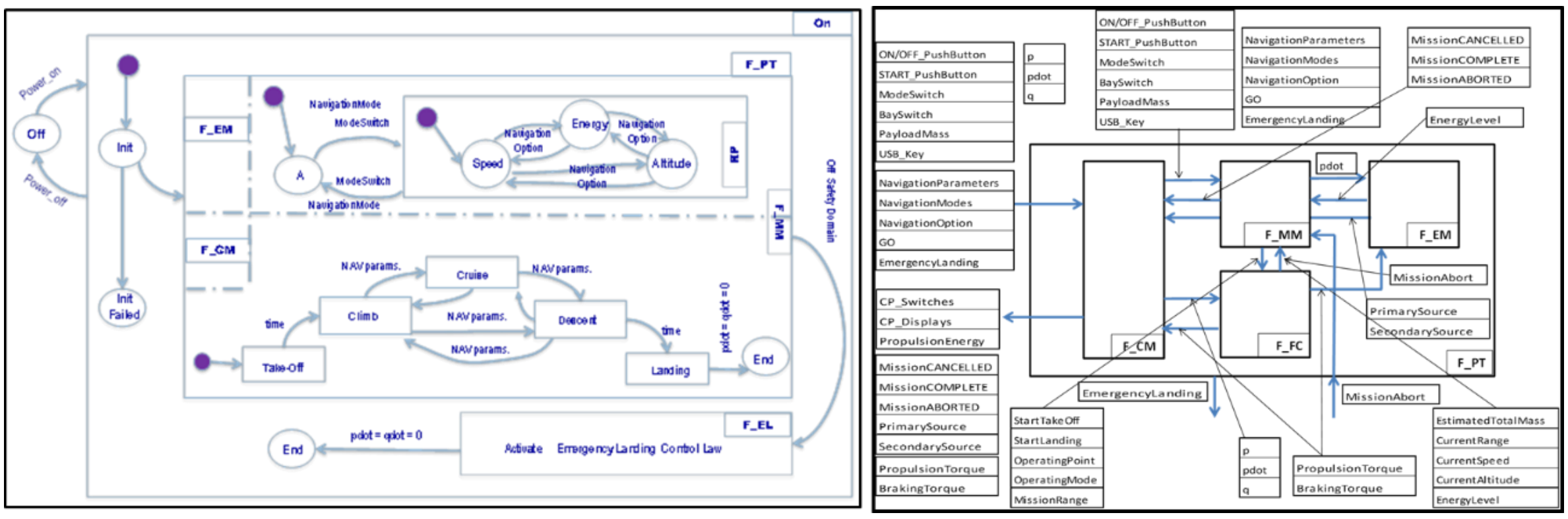

*Figure 12 – The nominal behavior. Left: the phase/mode/option Statechart diagram. Right: the functional flows exchanged between the core functions.*

Figure 12 shows how µXAV features *switched* control[35]. There are three sources of regulation mode switching:

1. Mission phase sequencing,
2. Navigation commands that change the reference values (emitted by the ground station, RP mode),
3. Failure modes that trigger fault-tolerance recoveries.

Figure 13 presents the overall control structure analyzed with STPA [Lev11]. For clarity reasons, it does not contain the FDIR loops.

[35] Non-linear control, both on the plant side (non-linear ODEs) and on the controller side (e.g. saturations).

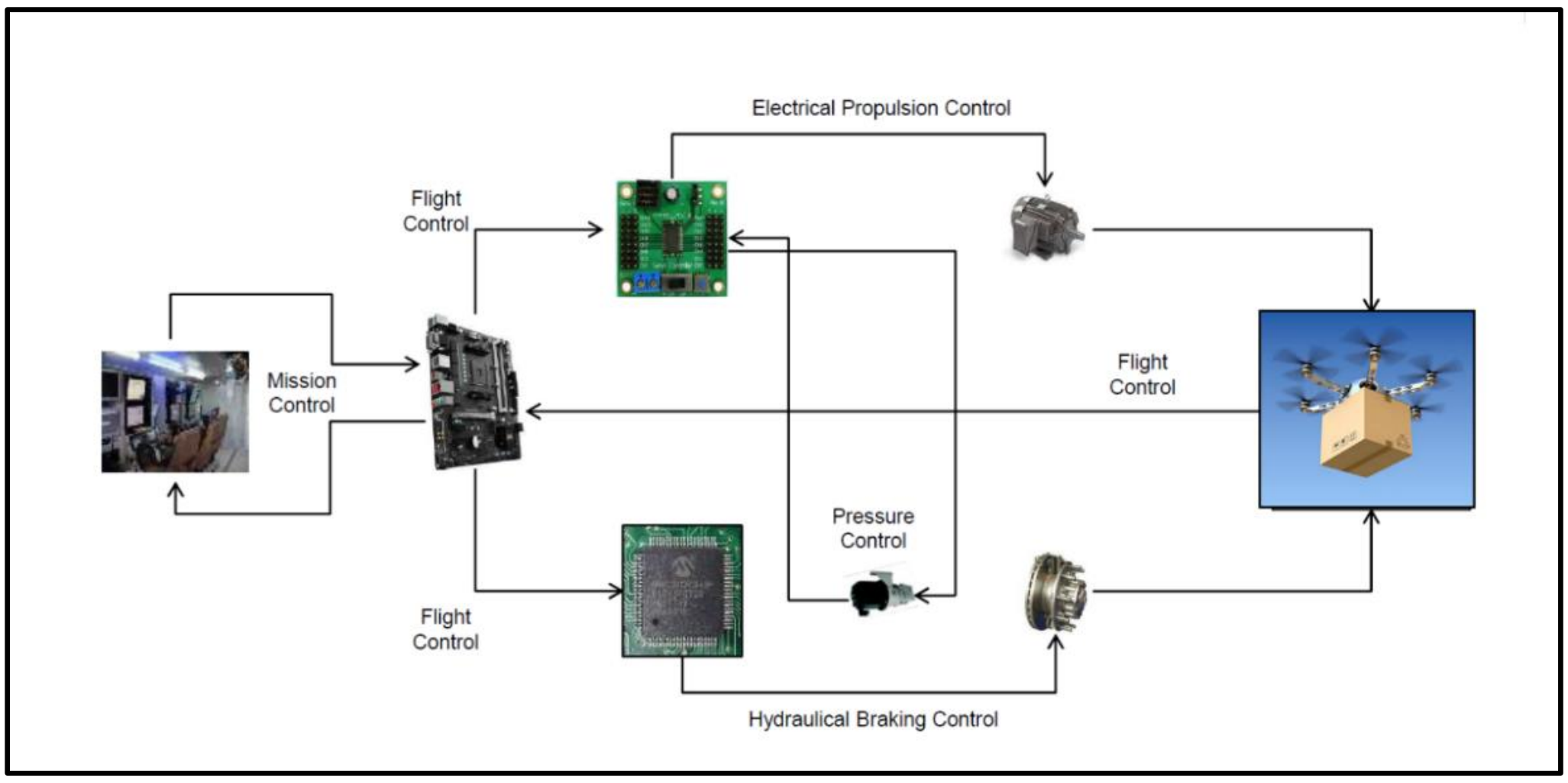

*Figure 13 – The control loop structure analyzed with STPA – Safety is addressed primarily as a problem of controllability under disturbances and uncertainties. Failure modes and failure propagations are regarded as one in many causes of unsafe control actions.*

The logical architecture and some aspects of the physical architecture are represented in figure 14. The four systems (MMS, CMS, EPS, HBS) have dual-channel computing and communication resources to be compliant with the “no single cause catastrophic failure” CS25.1309 requirement.

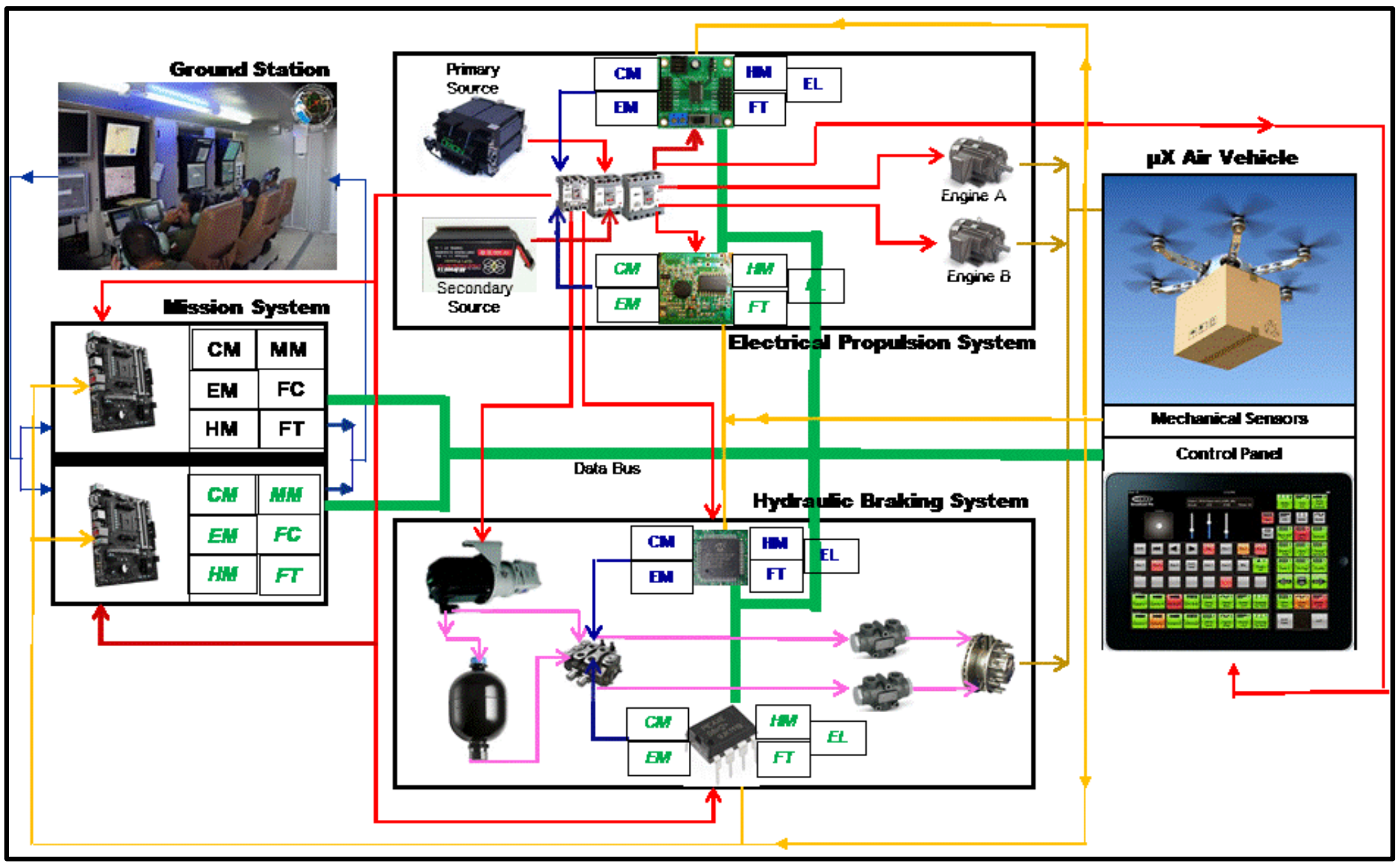

*Figure 14 – the µXAV duplex multi-system architecture*

## 3.9 The safety viewpoint

The air vehicle level Functional Hazard Analysis (AFHA), conformant to ARP4961 template, is available in the safety assessment folder of [RES18]. The µΩ project has scheduled completion of the ARP4961-compliant safety assessment process for the µXAV, i.e. completion of the PASA[36], of the four PSSAs[37], and of DAL[38] assignment. In addition, the control-oriented view of risk analysis and functional safety specification is experimented (see the STPA-related activities in “work in progress”).

[36] Preliminary Aircraft Safety Assessment
[37] Preliminary System Safety Assessment

The main functional safety properties are the separation envelopes at take-off and landing to protect the field operator. There are also flight-phase-dependent pitch envelopes and error bounds assigned to regulation. These properties are contextual invariants.

From fault tolerance by architectural mitigation standpoint (redundancy management), the main property is “no single point failure”. We use invariant-based design to craft with maximal independence the *parallel product* between mission control and FDIR control.

## 3.10 The security viewpoint

The µXAV system is open to digital user interactions. Wireless communications with the ground station and the control panel USB port are the two main vulnerabilities, with maintenance procedures. Security was out of the scope of the 2015-2018 use case. It is now in the scope of the µΩ project at THALES. In particular, we intend to explore invariant-based design supported by contracts that formulate the safety and security properties in assume-guarantee style. A twofold formulation of these contracts is under development: informal at Capella level, formal at Simulink, source code and hardware level.

Regarding the security aspects, a secured gateway in charge of isolating the inner communication domain (TSN[39] LAN) from the outer communication domain (wireless LTE/5G data-link with DTN[40] capabilities), is being developed in the CPS4EU project. This gateway is in charge of protecting the drone against spoofing, violation of mission-data integrity and confidentiality, denial of services, etc.

## 3.11 The fault-tolerance viewpoint

We consider fault-tolerance as the pivotal discipline of safety engineering. We deem it would deserve more specific assurance goals in development assurance standards, and the support of more powerful engineering techniques at design time. In the µΩ open research program, we would like to explore the following two groups of questions. They are relevant at any scale of integration, from SoCs to CPS/CPSoS. The crucial link between FDIR detection coverage addressed here, and sufficient controllability conditions is explained in §4.3 (complementation principle).

1. **Detection**. We assume formal definition of the perturbation domains and of the uncertainty domains, parametric and epistemic (cf. §4.1 on set-based engineering). We assume digital-twin orientation, i.e. model-based development at system and software level. In addition, we assume uniform structural analysis over the digital and the physical: abstract interpretation of the models extracts the influence networks that connect the perturbation sources to the safety-critical controlled variables[41] identified with STPA. In this context, is it possible to design a hierarchy of functional and organic detectors provably *complete* w.r.t. formally defined sufficient controllability conditions? For each local detector associated to a containment region, is it possible to characterize its detection coverage? In particular, what is Ω-probabilistic quantification of its false negatives?
2. **Recovery.** Latency in detection and reconfiguration leads to widening of error and failure propagation, until containment regions stop it. Latency also leads to greater deviations from the nominal behavior, possibly up to a point where pulling dynamics back in the controllability domain becomes unfeasible (controllability escapes). How to analyze the timing aspects of failure propagation? How to prove that transitioning to recovery *always* occur in conditions where sufficient controllability remains ensured, or else, that such recovery-failure conditions are quantified as sufficiently remote?

[38] Development Assurance Level
[39] Time-Sensitive Networks (deterministic Ethernet). LAN: Local Area Network
[40] Delay Tolerant Network
[41] More precisely, an over-approximation thereof. Regarding substantiation of separation and independence hypotheses, it is conservative.

Certification of autonomy motivates this formal approach to fault tolerance. There is a need of Ω-confidence on the transition between HPC and HAC[42] in the so-called “simplex” fault tolerance architectural pattern [Wan13]. This pattern is commonly mentioned as candidate safety-net solution for brittle AI-based vision functions relying on DNNs.

## 3.13 The certification viewpoint

[µCS17] contains a projection and adaptation to µXAV of the 1500-page CS25 regulation [EAS20]. The µCS25 requirements address the platform and the systems. Table 2 and figure 14 give examples of the adapted requirements. These properties are verification objectives for the set-based methods of section 4.

| Platform | Installation | | System |
|---|---|---|---|
| (CS 25.21) Centre of Gravity Variability Compliance Proofs | (CS 25.1301) Functions and Installations | PERFORMANCE | (CS 25.101) Performance General – Compliance in Still Air Conditions |
| (CS 25.23) Load Distribution Limits | (CS 25.1309) Equipment, Systems and Installations | | (CS 25.103) Stall Speed |
| (CS 25.25) Weight Limits | (CS 25.1310) Power Source Capacity and Distribution | | (CS 25.105) Take-Off |
| | (CS 25.1322) Instrument Installation: (Remote) Crew Alerting | | (CS 25.107) Take-Off Speeds |
| | (CS 25.1351) Electrical Systems and Equipments: General | | (CS 25.111) Take-Off Path |
| | (CS 25.1435) Hydraulic Systems | | (CS 25.121) Climb: One Engine Inoperative |
| | | | (CS 25.123) En-route: Flight Paths |
| | | | (CS 25.125) Landing |
| | | CONTROL | (CS 25.171) Stability: General |
| | | | (CS 25.181) Dynamic Stability |
| | | | (CS 25.237) Wind Velocities |
| | | | (CS 25.253) High-Speed Characteristics |
| | | | (CS 25.341) Gust and Turbulence Loads |
| | | | (CS 25.415) Ground Gust Conditions |
| | | | (CS 25.562)Emergency Landing Dynamic Conditions |

*Table 2 – Sample of the 30+ certification requirements transposable to the drone*

[42] High Performance Controller / High Assurance Controller

**(CS 25.117) Climb: General**

*Compliance with the requirements of CS 25.119 and 25.121 must be shown at each weight, altitude, and ambient temperature within the operational limits established for the aeroplane and with the most unfavourable centre of gravity for each configuration.*

Compliance with the requirements of µCS 25.121 must be shown at each weight, altitude, within the operational limits established for the drone and with the most unfavourable centre of gravity for each configuration.

*Figure 15 – Example of transposition of a CS25 regulatory requirement into a µ-regulatory requirement. Reference to CS 25.119 has been suppressed because CS 25.119 is not applicable to a drone.*

The µΩ project remains concerned by research on certification. [Sap18] describes a safety assurance process mapped to a model-based CPS engineering process. The geometric approach to dynamics presented fig. 7 and §4.2 and the controllability approach to safety underpin the assurance rationale. We plan to experiment these processes, supplemented by STPA, contract-based formulation of specifications, and safety cases.

## 3.14 CPSoS Extensions

Besides adding the security concerns we also need to add some system of systems aspects. While keeping the 2D only dynamics, in a first stage, we plan to consider fleets of coordinated µXAVs, equipped with ADS-B[43]-like board-board communications, and submitted to µ-ATM/ATC air separation and traffic regulation constraints.

## 3.15 Work in progress

We have extended Capella [Cap20] to support STPA in two modes:

- “stand-alone”, i.e. independently of any representation of the system of interest,
- “mapped” to Capella modelling of the system of interest.

We are experimenting this extension in both modes, starting with the former, then mapping and tuning the analysis with the latter. The first objective is to compare the identified risks with that of the baseline FHA [µFH17], and to assess STPA as a means of getting more complete safety requirements. Another goal is to contribute a control-oriented safety case, supplemented by contract-based development, and compositional control development (cf. §4.1).

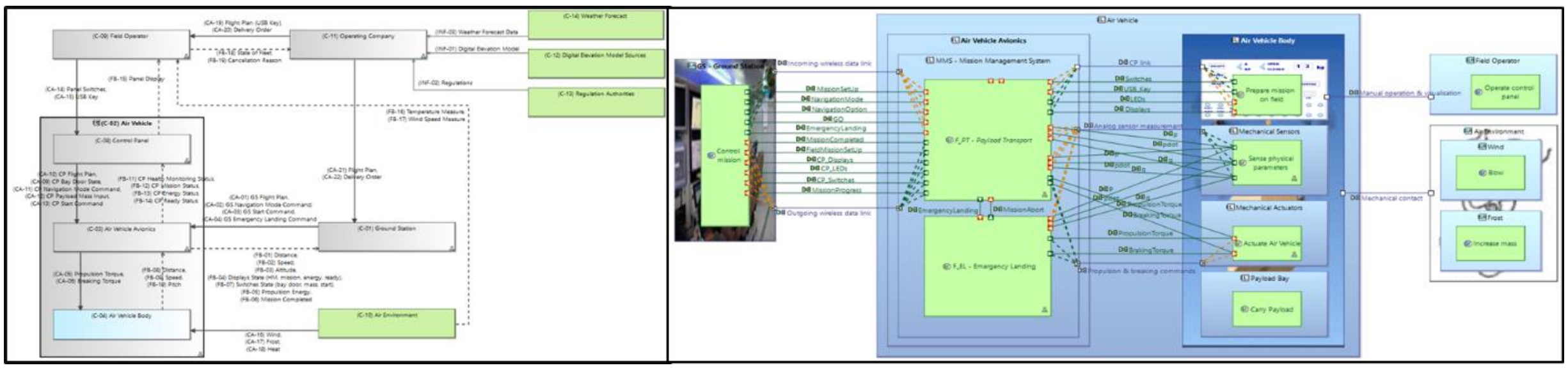

*Figure 16 – Left: Capella view of the top-level control structure analyzed with STPA. Right: Capella view of the top-level logical architecture, the items of which are possibly associated to that of the STPA views (e.g. Control Actions (CA), Feedbacks (FB), Unsafe Control Actions (UCA) etc.).*

[43] Automatic Dependent Surveillance – Broadcast.

We plan to experiment with STPA for any kind *controlled invariants*: safety, security, performance, usability, low power, etc.

Full-fledged Simulink behavioral modeling is planned (digital twin orientation). Figure 15 gives an example of formal verification of a digital-physical control property of the µXAV: speed and altitude reference capture and disturbance rejection using set-based verification (hybrid system model-checking [Sah17]).

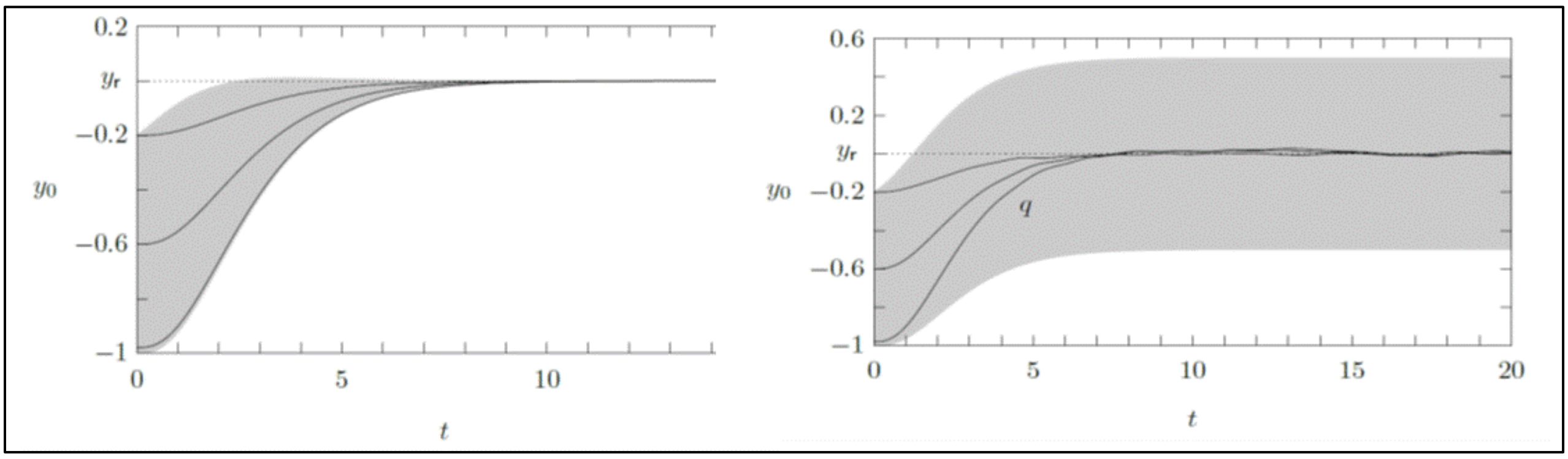

*Figure 17: Flow* formal verification of a dynamic regime of the µXAV, without wind perturbation (left), with wind perturbation (right). The grey area is the flow-pipe, i.e. the set of all possible behaviors given a set of initial conditions. Three simulation traces are superimposed on the flow pipe: two limit scenarios near the edge, and one "typical" scenario at the center of the usage domain.*

Finally, we intend to generate C code from the Simulink specification models, and model-based deploy the distributed implementation over the six computation units connected by a TSN network.

## 3.16 Engineering methods and tools

Support of end-to-end, top-down and bottom-up, informal and formal, cost-effective development of certifiable safety critical CPS/CPSoS is our main research goal. We build on the ARCADIA/Capella asset, interfaced with behavioral modelers and verification tools, enriched along the lines proposed in the next section. Certifiable AI-based autonomy, IoT systems and migration the cloud of highest criticality command and control are our main application drivers.

We intend to evolve the µΩ use case to support as many tool-oriented or application-oriented research explorations as possible along theses lines.

## 3.17 Synopsis

Safety critical certifiable embedded multi-system of the aeronautic domain at the origin, µXAV renamed µΩ is evolving to be compatible with experiments on a wider scope: cybersecurity, system of systems, certifiable AI machine learning, product-lines, set-based engineering, and interface with 3D-4 D physical engineering. However, engineering of trust and substantiation of trustworthiness, remain the main invariants.

# 4. Research Perspectives

In this section, we point to academic research we deem promising to overcome the challenges mentioned previously. It is by no means a state of the art. It is a selection of results to be applied to the µΩ experiment.

The first group of work (section 4.1) addresses what we consider insufficient generalization when specifying, i.e. missing logical regularization in system operational and functional design. We advocate that a generalization-oriented mindset is a key contributor to prevent the so-called "emergent properties" that cause high system and software IVVQ costs, and rework extra costs. Set-based methods, compositionality supported by contracts, and i*nvariant-based design*, are the three enablers we propose conjointly to master digital-physical behavioral complexity on large scales. In the end, it consists in lifting the formal approach now standard for hardware and software, up to the level of hybrid systems[44].

The second group of work (section 4.2) introduces a "new"[45] verification approach based on *geometry of dynamics.* We position it as an additional means to verify invariant-based design. The purpose is to add some *visual* engineering to the generalization policy enforced by invariant-based design. We want to engineer the *extensional* side of invariants, in addition to their *intensional*[46] definitions. The first group of work addresses the intensional, structured into assume-guarantee contracts [Sao18a], [Sao18b]. This second group strives to provide means to see the "shape" of the invariants, the structured tessellation they imprint on behavioral spaces. We sketch the potential impact of this approach on development assurance, more specifically, on what we regard as a case of misconceived dissimilarity requirement.

How to perform safety assessment on high fidelity behavioral models "in the large" is the point addressed by the third and last group (section 4.3). Engineering, on the deterministic side, ensures that the governed digital-physical stays enclosed by green-amber boundaries, or by amber-red boundaries, provided disturbance boundedness assumptions are not violated. When the scenarios violate the assume part of the controllability contracts, the dynamics may enter some incident or accident regions of the behavioral space. Engineering has then to switch to probabilistic analysis to "weigh the quantity" of entries into the red, and to prove that it is below the regulatory levels.

From a game-theoretic perspective, CPS safety engineering has first to prove that control always wins when the adversaries play within the disturbance mitigation domain. Second, it must prove that a win by the adversaries is sufficiently unlikely to be acceptable. Both are fundamental. Current industrial practice splits the qualitative controllability analysis and the quantitative controllability-escape analysis respectively between system engineering (SE) and safety assessment (SA). Regarding IVVQ cost reduction, or containment, one would reach the Ω-maturity level if *minimal* yet *provably sufficient* behavioral exploration of the high fidelity models could deliver the deterministic and the stochastic proofs *at the same time*, by the same intensive behavioral exploration[47].

S*calability* of modelling and analysis is a major concern of this section. We have in mind autonomous vehicles, collaborative robots, and digital-physical OODA loops that would span over CPS/CPSoS such as cloud-controlled railway infrastructures, or crisis management systems. However, the referenced papers all rely on state representation of dynamical systems and iterated transition maps. Distribution-induced asynchrony must remain bounded into rounds [Bro98] or into any variant of virtual synchrony [Bir10] to remain compatible with this theoretical setting.

[44] Hybrid system theory (e.g. [Tab10]) is the overall theoretical background we have selected to found our CPS Ω-engineering.
[45] It traces back to Poincare 1888
[46] i.e. logical, symbolic. Following [Sim96], the Intention -> Intension -> Extension parallels the Ends -> Laws -> Means categories of Herbert Simon.
[47] Intensive simulation campaigns on models (specification-based development) or testing DOEs on legacy and COTS (integration-based development). In both cases, first requirement-based testing, then adversarial testing.

In the three 4.x sections, we assume CPS/CPSoS to be modelled in a way compatible with [Tab10], [Sao18a], [Sao18b] for the governed digital-physical: a transition *relation*[48] over finite and infinite state variables, dependent on internal and external inputs, and structured into interacting nodes. This setting enables specification and refinement of *behavioral constraints*[49]. They enable leaving unspecified any behavior-solution to the constraints. A fortiori, they leave any implementation undefined. Addressing control specification first by defining behavioral *envelopes* is key for the early stages of system engineering, especially on large projects where many specification and implementation tiers are involved.

Section 4 is speculative. It proposes and substantiates some insights and orientations. They are rooted in three decades of system and software engineering practice in interaction with academic research teams. To remain concrete, we instantiate these prospective considerations on the μΩ-use case.

## 4.1 Set-based compositional system engineering

Set-based in this section does not mean interval-based, as common in the constraint-solving community. “Set-based” means “set-valued” [Aub94], to contrast with “point-valued” modeling and analysis of current system engineering. The sets may be represented by any means: intervals, support functions, affine forms, polyhedrons, polytopes, zonotopes, simplicial complexes, etc. A point-valued execution is a simulation trace on a host, or a testing trace on the final product. Abstract interpretation or software model checking perform set-based execution of software. On system models, set-valued execution is for instance flow pipe and reachable state space computation [Fre11], [FH06], [Ale16].

Besides formal specification, synthesis, and verification of the governed digital-physical, set-based methods are also relevant for 0D platform design and trade space exploration [Seb20], [CVT20]. Set-based engineering can uniformly encompass platform and system, 0D and 1D, as illustrated on figure 9.

“Compositional” means enabling modularity in specification, implementation, verification and validation. The formal techniques now standard for hardware and software development are being lifted to the system level, physics and control inclusive: composable specifications, predicate abstraction, specification refinement, implementation refinement, invariants, and assume-guarantee contracts. An end-to-end uniform formal approach to system, 1D physics, software and hardware development is emerging.

[48] More general than a transition function
[49] Differential inequalities, differential inclusions, higher-order automata, etc.

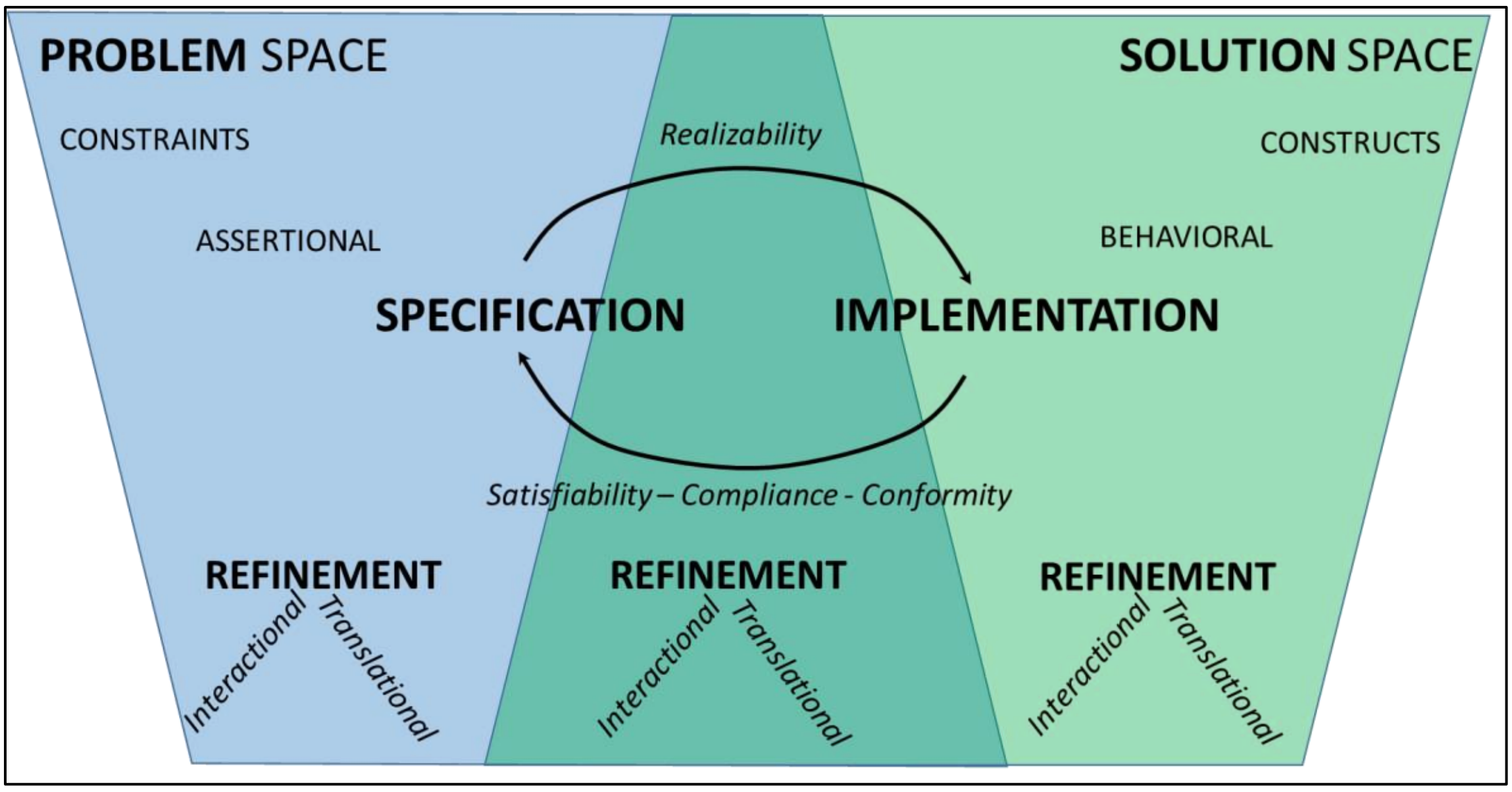

*Figure 18:* From "Construct-by-Correction" engineering, to "Correct-by-Construction" engineering: abstraction-refinement of specifications followed by synthesis of constructs *from* constraints, of actions *from* assertions, i.e. verified-*while*-developed. The blue-green merge area denotes fusion (possibly confusion as well) between needs and means, between problems and solutions. Control specifications are often exclusively behavioral, in which case refinement into implementation is primarily translational by nature.

### 4.1.1 Why set-based methods?

Current engineering probes specification and implementation correctness. It specifies and verifies on a case-by-case basis, by defining a set of typical and limit scenarios from which the test cases are defined. No wonder unintended behaviors may "emerge" in operation after costly intensive testing campaigns.

"Emergence" emerged in statistical mechanics, in physics of phase transitions. It has spread in the AI multi-agent system community, and in the system engineering community. In statistical mechanics, it denotes emergence of order out of disorder. Not of disorder out of order.

Set-based engineering addresses *spaces*:

- Variability domains, configuration spaces for product lines and platform-based development,
- Variability domains of components for integration-based development (e.g. IoT),
- Parameter spaces in platform and system sizing and optimization,
- Uncertainty domains on physical parameters and laws,
- Design spaces, i.e. solution manifolds constraint problems,
- Behavioral spaces, i.e. hybrid system reachable state spaces.

The goal of set-based methods is to capture the variants, invariants and constraints of the engineering problem, to compute the *feasibility domain,* and then to synthesize or to "generate and select" some solutions within it. This process ensures correct-by-construction designs, without excluding manual design[50], and without excluding some impact on "emergent properties" and IVVQ costs either.

### 4.1.2 Why contract-based engineering?

Contract is the pivotal notion that supports "decoupling" and modular development. For highly integrated control-intensive systems, modularization of design and verification is hard because of the twofold cyclic nature of control. Modular development of the governed digital-physical has appeared [Sao18a], [Sao18b]. Introduced in software engineering by B. Meyer for Eiffel contracts are spreading to system engineering (e.g. [Wil16]). [Ben18] contains an extensive bibliography. We follow [Bro98] for compositionality and abstraction-refinement of the distributed digital. We follow [Tab10], [Bel17], [Sao18a], [Sao18b], [Bog14] for compositionality and abstraction-refinement of the governed digital-physical.

[50] Manual but machine-assisted since verification is performed *while* designing.

### 4.1.3 What is invariant-based design?

In computer science, the way of mastering the infinitely many behaviors of programs has been to define state invariants, to recognize instructions as state transformers, and to reason by induction on transformation sequences. Floyd, Hoare, Dijkstra, Knuth, Milner, Cousot, to name a few, have opened that path. Today, development of large trustable software by means of mechanized proofs of invariants is an industrial reality for OT, and for complex software tools like compilers or operating systems [Ler16], [Bou20], [Kle10]. Thinking of specifications and implementations in terms of state transformations and invariants [Bac09] has proved its effectiveness for the digital[51].

Long before in physics[52] other messy situations of infinite behavioral complexity have also been resolved, factored, structured, by looking at the invariants[53] of selected groups of state transformations. Still even before, in mathematics, the very same approach had been introduced to sort out the relationships between the emerging non-Euclidian geometries. Klein, Lie, Killing, Noether, Poincare, Cartan, Weyl, to name a few, have been the forerunners that opened this path in geometry, geometry of dynamics, and mathematical physics.

The main prospective claim of this paper is the following: after the physical and the digital separately, the same approach will succeed to master the behavioral complexity of the governed distributed digital-physical. It will probably also need as much time as was needed in the two previous cases.

Control is amenable to set-based, invariant-based, and geometry-based methods [Bla15], [Won85]. Lyapunov functions is probably the most well-known historical example. In the piecewise affine case, scalability is that of matrix algebra. In theory at least, any dynamical system can be approximated, at any level of precision, in this setting. The convex non-linear case scales-up likewise. In the most irregular, singular and curved cases, one needs to resort to differential geometry and topology. Sensitivity to dimensionality limits tractability to partial studies; anyway, the valuable insights extracted locally still benefit more global behavioral analyses at higher levels of integration.

As control in OT is dominantly control of physics, and since physics has been geometerized, set-based and invariant-based methods for the governed physical have geometric foundations and visual interpretations. This motivates the next thread of research (§ 4.2), with potential positive side-effects on training and future development assurance.

### 4.1.4 μΩ-explorations

The set-based approach will be applied to the 0D and 1D engineering activities illustrated by figures 9 and 17. Contracts have been defined for safety and mission performance, at top level, system interface, function interface, and some component interfaces [RES18].

Invariant-based design has been applied at functional specification level, to enforce compositionality and symmetry. For instance, FDIR has been designed as uniformly "orthogonal" to the nominal modes. The invariants will be used to quotient the continuous dynamics [Tab10], using predicate abstraction

---

[5151] The hardware part of the digital; only on niches for the software part. Financial and legal impacts of bugs are significantly different between software and hardware.

[52] Electromagnetism and special relativity, general relativity, rational mechanics

[53] Named symmetries in physics

[Slo13], [Mov13], [Bou15], on the assume guarantee predicates of the contracts. FDIR detectors coincide with some of the contract predicates.

We plan to handle the finite-state quotiented phase spaces, and some of the vector fields, as explicit engineering artefacts accessible from Capella. The structure phase spaces is made of sums, products, and gluing of invariant-based behavioral cells. Each cell is an equivalence class, at system level, that mirrors the similar notion in software or hardware testing. This will manifest the behavioral space *architecture* of the µXAV, on which testing coverage sufficiency will be demonstrated.

The planned typical set-based-IVVQ use case is incremental recertification: given a motor or battery pack modification, based on analysis of the preserved invariants, and on contract violations, a minimal and *provably sufficient* re-verification campaign will be computed, applied, and validated on a "digital twin" mock-up of the µXAV.

## 4.2 A geometric approach to physical dynamics

When Henri Poincare tackled the 3-body problem by the end of the 1880s, he discovered that it was impossible to exhibit an analytical formulation of the celestial bodies' trajectories. To decide the asymptotic stability of the system in spite of this impossibility, he managed to extract a new kind of information from the equations: properties of the *geometrical locus* of the trajectories. These trajectories were not located in the ordinary 3D physical space. They evolved in a higher dimensional space where the *global state* of the 3-body system was represented at instant t by a single 18D-point, which he managed to reduce to a 6D-point with simplifying hypotheses and coordinate transformations.

Named *qualitative dynamics* this method has been the origin of major developments in mathematics, and has found applications in physics and control theory. It is only recently that some computational geometry libraries, originally developed for 2D-3D applications, have proposed a comprehensive set of algebraic topology and differential topology functions in *any* finite dimensional space, opening thereby new perspectives for verification of the physical, and of the governed physical.

It happens that with these new libraries it should become possible, for low dimensional[54] hybrid systems, to compute a geometric model of the *edge* of their nD phase space. Such a model would help solving three important problems of system safety engineering.

### 4.2.1 Decision manifolds

Return on experience has shown that designing sensitive, but not too sensitive, safety monitors (the D of FDIR) is tricky on complex controlled physics like flight mechanics, thermo-fluidics or cobotics. One may never find the good tuning of logical conditions and threshold parameters to reduce *simultaneously* the false negatives and the false positives. It suggests that purely analytical and logical methods to design the decision-observers, and simulation-based verification, may not be powerful enough in the most complex cases. These complex cases are the very cases where correctness of the safety monitors is the most critical. Safety nets of AI-ML functions and simplex architectures [Wan13] are within this scope.

If the phase space is n-dimensional, the decision frontiers or "zero-crossing surfaces" are (n-1)D manifolds[55]. Complexity of the shape of these decision frontiers grows very quickly with the number of variables and with the nature of the mathematical functions used in the decision predicates. Could geometrical engineering help verify observers by an extensional analysis of their intensional definition?

[54] $1< n < 10^{+}$. Some of the needed algorithms are sensitive to the curse of dimensionality.

[55] By default, on most standard cases. For instance, linear separation hyperplanes for one-parameter thresholding. In "degenerated" or singular situations, the dimensionality of the separation manifold may be lower than (n-1).

### 4.2.2 Safety properties

Safety properties are contextual invariants: “always within such domain in such conditions”. They define barriers. Control, especially FDIR, compensates for the disturbances and pulls the dynamics back into the nominal domain to prevent escapes beyond the barriers. Viewed geometrically, safety properties are *holes* in the phase space, no-dynamic regions. These holes are in contact with the edge of the reachable state space. If one could compute its geometrical representation, one could analyze its intersection or inclusion w.r.t. decision manifolds.

### 4.2.3 Behavioral coverage measures for the physical

As seen in section 2, requirement coverage of testing does not suffice to prevent the unfortunate “emergent properties”. For the physical, there is no structural coverage like for the digital. Geometrical modeling of the edge of phase spaces might open to computing *the volume* of behavioral space regions, in particular when tessellated by invariant-defined cells. Using scoring functions[56] and level set techniques in addition to volume computation, behavioral spaces of the physical might become objects that are *structured*, *measured*, and represented into engineering artefacts (e.g. smoothed simplicial complexes). Consequently, they would become amenable to *behavioral coverage* analysis.

Then, one could use this set-based geometric engineering to justify the positioning of the test trajectories[57], i.e. to justify the *sufficiency* of IVV test plans by substantiating:

- how the trajectories visit the cells of the structured reachable state space,
- how they are centered (typical scenarios) or “boarder-line” (limit scenarios, corner case and edge case scenarios) w.r.t to the edge of the behavioral cells,
- how the volume of the *unexplored parts*[58] is acceptably “negligible” in the given context of development assurance.

### 4.2.4 How to do it?

Reachable state spaces may have very complicated structures of inter-woven trajectory bundles. $XX^{th}$ century mathematics has classified these structures. Non-linear control engineering has drawn from this theoretical background. The nD-building-blocks of phase spaces have intuitive interpretations: attraction basins, repeller basins, saddle points etc. Our view of CPS/CPSoS Ω-engineering would involve computation of invariant sets and topological invariants on *architected* reachable state spaces for the governed physical.

For instance, computation of the first Betti numbers on the simplicial complex of the edge of a reachable state space can detect coalescence of some holes, in other words it can detect the “vanishing” of some no-dynamic regions, the violation of some safety invariants. Another example is the computation of maximal invariant sets in a given region using the Wazewski principle [FGM15], [FGM16]. Roughly speaking, one probes a region of the vector field structure expected to be green or amber with a “closed hull”. One wants to verify that no controllability escape may happen within this region. One analyzes the vector field, i.e. the trajectories, at the hull’s boundary. If everywhere the vector field[59] enters the boundary, then differential topology theorems ensure that its interior contains a maximal invariant set enclosing an equilibrium point or a limit cycle. In other words, it contains a stable regime bounded by the hull, so devoid of controllability escapes.

[56] For instance hazard scores, utility scores, performance scores etc.
[57] Vectors of time series are point-valued trajectories in phase spaces.
[58] Analogous to unactivated code in software (the dead code elimination issue).
[59] Defined by the state-space representation model of the CPS/CPSoS

At short-term and mid-term we plan to focus primarily on the linear case, and on the "skin", the edge, of accessibility spaces. We would like to "mesh these skins", by means of simplicial complexes. The general idea is somehow analogous to reverse engineering of buildings and old mechanical parts manufactured before the advent of CAD-design: first, a laser beam scans the existing 3D object; then 3D computational geometry algorithms reconstruct a model of the scanned surface, a model of the "skin" of the legacy mechanical object. The surface, the edge, is learned or inferred from the point database.

In our case, the nD phase space replaces the ordinary 3D space of mechanical engineering. The anticipated method would first compute a tight bracketing of the edge by simultaneous use of outer- and *inner*- approximation of the reachable state space [Gou17], [Gou19], [Gou20]. This geometrical interval brackets the edge but its true location within is still unknown. The second step would consist in massive generation of short trajectory segments starting from the edge of the inner-approximation. This step is the analogue of scanning in the 3D reverse engineering example. The third step would use nD-reconstruction of a simplicial complex of the higher dimensional "skin" [Boi18], [GUD20], [Cha17]. The fourth step consists in using such geometrical objects to solve the three aforementioned problems.

### 4.2.5 Certification motivations

For some of the most recent aircraft, certification Authorities have issued documents[60] that formulate concerns on the growing complexity of flight control laws, and on the consequent risks of single cause catastrophic failure at *specification* level. They deemed necessary to ask aircraft manufacturers to propose and implement additional means to mitigate these new risks. Doing so, the Authorities acknowledge that the safety assurance standards become too weak for the most complex control specifications. One means the Authorities have suggested as acceptable to mitigate this risk, is *2-version development* of the flight control laws specifications, by two *independent* teams.

We consider this orientation as symptomatic of the widening gap between the perceived power of current engineering and assurance, and what Authorities think would be necessary to address trustworthily the new classes of control complexity. Two-version development would more than double the development costs. For what safety benefit?

Consider the central problem of designing safety monitor decision manifolds, mentioned above and illustrated by figure 18, derived from figure 7a.

[60] Issue Papers

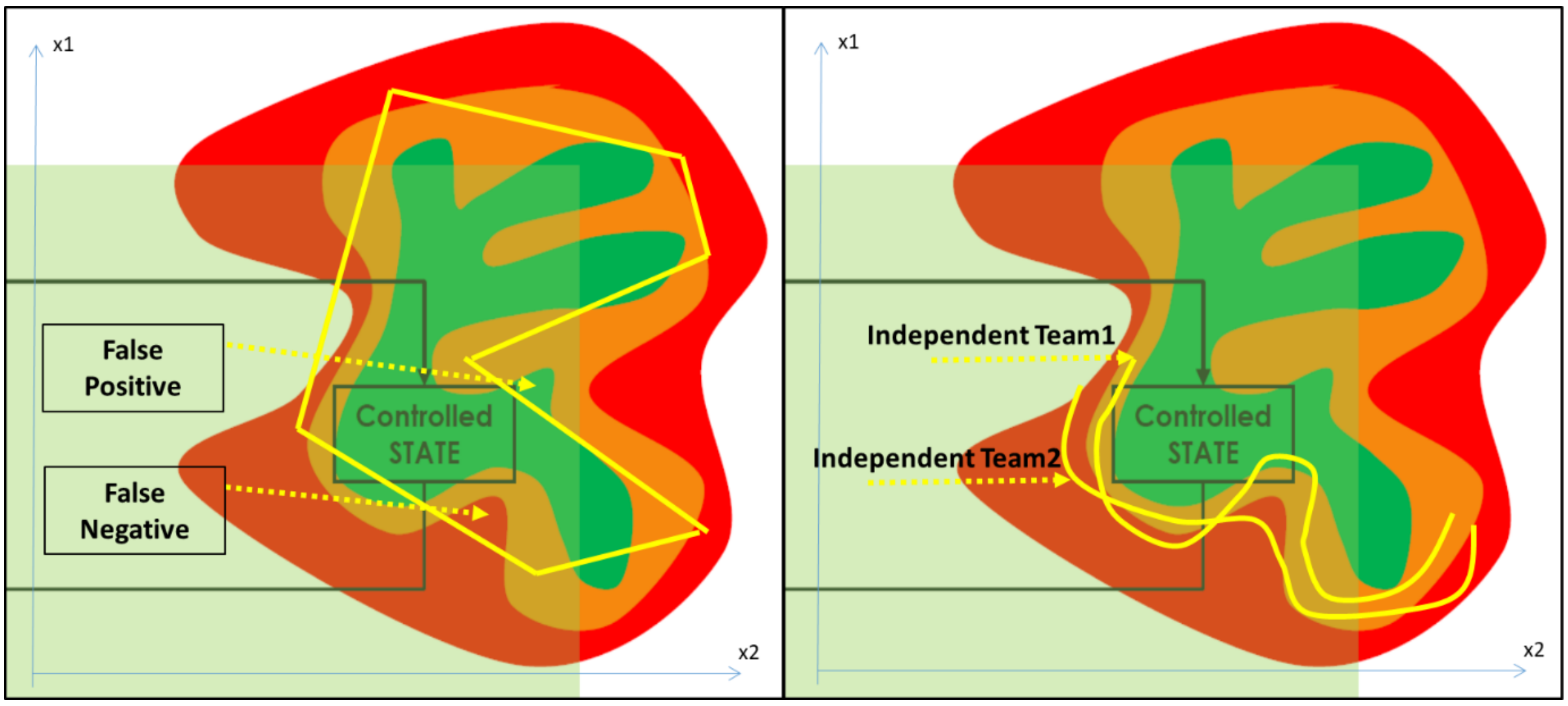

*Figure 19:* Design of a 1D-decision manifold (yellow lines) in a 2D phase space. The FDIR safety monitor must decide the green-amber frontier to switch to a back-up configuration before reaching the amber-red frontier. Fig18a: the system specification handles the decision problem in a linear setting. The green-amber frontier and the amber-red frontier of the behavioral space represent "ground-truth" of the observation-decision problem. The dashed yellow arrows point to two incorrect decision regions, among a few others. Fig18b: assume two teams have developed "independently" the same safety monitor. Both have chosen non-linear methods to define the decision predicates. The two yellow curves represent part of the two results. What is the decision in case of disagreement between the two channels? How to consider "independent" these two curves that strive to conform to the same complicated "behavioral shape"?

We argue that geometric complexity of dynamics in higher dimensions is likely to constitute a common root cause of specification error that may ruin any attempt to mitigate this risk by two dissimilar developments. The orientation of the μΩ project is to introduce new verification means to augment trust on a *unique* specification. We would like to show:

1. the safety benefits of resorting to new symbolic control verification techniques [Bel17], [Sao18a], [Sao18b] and new geometric verification techniques [FGM15], [FGM16], in *addition* to standard testing campaigns, supplemented by adversarial testing,

2. how behavioral complexity can be controlled and *architected* with these new set-based and geometrical approaches, at specification and implementation levels,

3. how this geometric representation of behavioral spaces can help implement Bret Victor's "immediate connection" principle[61]. We regard it as beneficial not only to education or corporate traning, but to safety engineering and development assurance alike [Vic12].

### 4.2.6 μΩ-explorations

The safety and performance contracts of the Air Vehicle have been defined geometrically (set-based approach). The global behavioral space will be architected, explored, and abstracted as a geometric object. Geometrical analysis will be performed on the detection domain of the following FDIR safety monitors:

- flight control loss or malfunctioning,
- batterie and motor loss or malfunctioning,
- excessive energy consumption.

The E (energy) navigation mode is an optimal control mode to illustrate geometrically interpreted "verified-by-design" control law synthesis. Systematic generation of invariants for augmented observability and correct-by-design detectors is scheduled (parity space methods [Fri17]).

[61] Possibly not so "immediate" when n > 3, but even so, assurance benefits are expected

## 4.3 Deterministic-probabilistic co-verification on high-fidelity models

We outline an alternative approach to safety assessment in which we address the third open problem (cf. §2.3). We refer to the safety standards of the aeronautic domain [ED79] [ED135], but its applicability extends to other domains like railway, automotive, space, nuclear etc. This approach draws from the two previous sets of foundations and emerging techniques: predicate abstraction to reduce hybrid systems to composable finite-state transition systems, and the geometric view of behavioral spaces (green/amber/red regions).

We first make explicit the underlying rationale. The safety assurance rationales are often implicit in the standards, and hidden behind process-oriented presentations, if not silenced on the most difficult aspects. In this paper, we are dealing with *the gaps* and *the engineering barriers*; our point is precisely to make explicit a few difficulties that hamper clarity of the safety standards and boundedness of negotiations between applicants and Authorities. Doing so, our purpose is *not* to cast suspicion on past and current certification practices. Current standards record the *best* practices negotiated between applicants and Authorities. Our point is to cast foundational light on a few core problems we deem are becoming tractable and candidate for mid- to long- term evolution of the negotiated recommended practices.

We revisit the articulation between deterministic and probabilistic engineering in safety assessment, with *scalability* and *accuracy* as primary concerns. The method outlined beneath aims at addressing any scale of integration, from SoCs[62] to fog computing and CPSoS, as discussed in section 1.

### 4.3.1 Rationale

- "There is no zero-risk system" (Cl1),
- "Future systems will be too complex to be addressed by deterministic approaches" (Cl2),
- "Formal methods do not scale-up and/or are not affordable" (Cl3).

Most often, these claims are used altogether to substantiate an inevitable all-encompassing statistical approach to IVVQ. The booming of AI-ML, i.e. of programming by statistical estimation of functions, has reinforced this trend.

We claim that scalability of *valid* probabilistic calculations would be scientifically and technologically as hard as lifting formal methods to address these new scales of system integration and behavioral compexity. Reliability approaches are attractive in system certification because of acceptance of simplistic Boolean models (combinational or sequential) to represent physical-digital behaviors of extreme complexity. *Trustworthy* probabilistic CPS/CPSoS analysis would require to start with the Piecewise Deterministic Markov Process view (PDMP) [Dav93], and then to justify soundness of conservative abstractions down to simpler finite-state models amenable to combinatorial probabilistic computation [Kwi11], [Bou08].

Following research in symbolic control for design and verification of hybrid systems, we support the lift of formal methods from software and hardware to the governed digital-physical. Model abstraction-refinement[63] provides means to combine probabilistic and deterministic behavioral analysis with *balanced model fidelity*. We do *not* oppose the deterministic to the probabilistic. We propose to combine them in a more intricate manner than practiced today, without the present split between system engineering models and safety assessment models. We point to the *unbalanced rigor*

---

[62] System on Chips

[63] More precisely the Counter-Example Generation Abstraction Refinement (CEGAR) approach

between the two engineering domains[64]. Would Authorities enforce assurance on probabilistic calculations on par with that enforced on software development and software tool qualification, the pan-statistical approach would be less attractive.

The underlying principles of the rationale are the following:

**Controllability Approach to Safety.** We follow the line of thought according to which *interaction failures* without component failures[65] are becoming a major risk on large-scale software-intensive, control-intensive distributed systems [Lev12]. It is a risk of functional safety *specification error* at system, multi-system, and system of systems level. Loss of control should precede loss of function in risk and causal analysis.

**Containment of Behavioral Complexity.** Maximal separation between tractable digital-physical influence networks is the traditional and primary solution to manage behavioral complexity. Structural analysis, based on abstract interpretation of models of the digital, provides scalable means to prove separation of influence cones. When *structural separation* is not possible, invariant-based design provides means of *logical separation*.

**No Inverse Dysfunctional Behavioral Analysis.** We have revisited the nature of FTA when used beyond its original semantics. FTA is a *structural* integrity-safety method. We keep FTA, in strict structural loss-semantics for DAL assignment in conformity with [ED135] and similar standards. We propose the *complementation principle* to address incompleteness of failure cause identification on complex CPS/CPSoS.

**Complementation principle.** It is a consequence of our rejection of the completeness claim on accident cause Boolean modeling[66]. We propose to shift the completeness objective from accident causal analysis to *decidable sufficient controllability conditions*. We propose to concentrate system engineering *and* safety assessment on mastery of the fault-tolerant controllability domain. This is unknown unknowns inclusive. Either we do not care because their effects are controllable; or we do not care because their effects violate the controllability contract. However, if some unknown unknown effect is undetected while pulling dynamics outside of the controllability domain… Completeness and correctness of the FDIR decision logics are critical anyway, but even more so with our proposed controllability-centric approach to quantification. For insight only and roughly speaking: the "weight" of escapes into red regions would be *over*-estimated by the "weight" of *all* behaviors minus that of the fault-tolerant controllability domain. Complementation means quantifying the adversaries' win by (1 – Pr("FDIR wins")) instead of direct computation on a claimed complete model of the entries to red regions. Could such an implicit-residual approach be well-founded, tractable in practice[67], and cost-effective on the large systems addressed in section 1?

[64] A number between 0 and 1 is scientific, trustworthy. Even more so, as understanding probability and stochastic process theory to assess validity of the calculations is hard. For instance, are the independence hypotheses made, or the distribution tail-dependence estimations, scrutinized by development assurance practitioners in a way apportioned to their critical role in validity of rare event quantification? [Gob13], [Mor16].

[65] In aeronautic wording, one could use the oxymoron "failure-less failure conditions", where the first "failure" abbreviates "component failure modes"

[66] This rejection is also motivated by value-for-money considerations [Led20].

[67] The quantified set of behaviors would be made of the nominal ones and of those generated by low-order (1 to 4) superposition of initiators of the SOTIF and Integrity domains (cf. fig. 6). The non-analyzed residual is gigantic from enumerative perspective, but rare, and of no-value embedded in the product.

**Deterministic-Probabilistic Co-Verification**

The rationale is to explore and measure the reachable state space of the fault-tolerant controllability domain, exhaustively wherever possible (cf. formal approaches of 4.1 and 4.2) and by testing anyway. First, standard requirement-based testing on the nominal and on the 1-disturbance cases. Second, on behalf of the revisited safety assessment, an adversarial testing campaign of the controllability contracts with 1-to-4 disturbances that would target and *count* the amber-red crossings. In addition to refutational[68] information (counter-examples), statistical estimation of failure events would be derived from the crossing-counts and from the biases in the input distributions (variance reduction methods [Gob13], [Mor16]). Such an approach would rely on concentration inequalities to estimate probabilities. It would use neither asymptotic results nor Boolean quantification.

**1-Development N-Verifications**

By verification we mean set-valued analysis (formal methods) or point-valued analysis (simulation and testing), both possibly deterministic (assertion satisfiability) or probabilistic (rareness estimation). In 4.2 we have questioned the relevance of dissimilar development to cope with single point failures at control specification level. More generally, our rationale favors uniqueness, seamlessness and digital continuity of a unique development flow. It promotes redundancy and dissimilarity for *verification*.

[Ler16], [Kle10], [Bou20] and the considerations of 4.1 and 4.2 motivate our stance. Regarding some of the sophisticated transformations involved in development processes[69], one can now compile with unprecedented trust a Domain Specific Language (DSL) dedicated to control, down to assembly code. High assurance on interactional refinements of the governed digital-physical remains an issue. When one decomposes a behavioral entity into interacting sub-entities, one introduces a combinatorial parallel product that constitutes a combinatorial source potentially hard to verify with proved sufficient coverage.

### 4.3.2 Co-verification process

The 1-Development N-Verification policy requires from verification techniques to scale on the development artefacts since there are only development models to verify. It bans the notion of verification model. CPS/CPSoS development may be dominantly integration-based, reusing large legacy parts and components off the shelf. Set-based, contract-based and invariant-based design apply preferably to top-down specification-based development, and even more preferably to model-based development. However, we need to apply our approach also, if not primarily, to bottom-up integration-based development.

Reverse engineering of specification traits from black-box extensive testing of software, hardware or system reused components, is a necessity [Bol10], [Isb15], [Aar15] (e.g. learning of temporal logic formulas or finite-state transition systems). It is a prior to bottom-up *contract-matching*. When partial specification learning from logs or DOEs is not possible, manual reverse contract-based specification is inescapable.

Infinite state verification does not scale, be it set-based methods or massive adversarial testing. It can address complex use cases, and is of critical importance to verify invariants on the governed digital physical. However, to address the global level of CPS/CPSoS models, one must faithfully reduce or "summarize" the infinite-state component models into finite-state ones. This reduction step "discards" the continuous physical dynamics. It keeps only the evolution of the observational predicates defined for finite-state reduction. This reduction is observational, viewpoint-based. For a given component

68 Falsified assertions of the contracts (assumes or guarantees), or of any kind of test oracle.
69 Named translational refinements in figure 18.

system one may need to perform as many finite-state reductions as needed by the properties to verify at global level (deterministic or probabilistic).

The envisioned deterministic-probabilistic co-verification process decomposes into four steps:

1. Infinite state verification on the high fidelity system engineering models (deterministic and probabilistic) [Ale16], [Bel17], [FH06], [Fre11], [Slo12],
2. Reduction of the infinite state parts to interacting (timed)[70] finite-state transition systems, [Bog14], [Bou15], [Mov13], [Slo13],
3. Possibly addition of safety assessment specific (timed) finite state transition systems when system engineering is model-free, or when reduction of high fidelity models by predicate abstraction is intractable,
4. Verification of the global (timed) finite state composition of the local finite state reductions (deterministic and probabilistic) [Kwi11].

**(Infinite - Finite) .vs. (Continuous-Discrete)**
Continuous variables, indexed by continuous or discrete time, are infinite-state. Finite-state variables are discrete. Discrete variables may be infinite-state: texts, strings, graphs, structured objects like situational awareness or mission planning objects.

**(Infinite - Finite) .vs. (Local - Global)**
Scalability of global verification requires reduction to finite state. Infinite state verification does not necessarily imply local verification. Global requires *partial*, composition of “résumés”, of viewpoint-abstractions. An infinite-state hybrid model may represent some mode-dependent physical processes that are distributed over a smart city or a railway infrastructure. Large spatial scale may remain compatible with infinite-state accessibility analysis. What matters is the size of the reachable state space, not geo-localization strech.

### 4.3.3 μΩ-explorations

A high-fidelity behavioral model, a “digital-twin”, is under development. An augmented version dedicated to verification will contain the observers that implement the “assumes” and the “guarantees” of the contracts and the failure modes detected by health monitoring and compensated by fault tolerance (controllability domain).

The main objective will be to carry out the three-stage, possibly four-stage approach:

1. Infinite-state deterministic-probabilistic verification of the system models separately: MMS, EPS, HBS and CMS,
2. Predicate abstraction of the four hybrid system models[71]. The observational viewpoints will use some environmental predicates (e.g. “WindBeyondLimits”), mission performance predicates (e.g. “CruisePrecisionOK”), failure mode predicates (e.g. “motorA_FAIL”) and the assume-guarantee predicates of FDIR contracts.
3. The four finite-state reductions will be composed into a global finite state reduction verified, deterministically (e.g. the FDIR reconfigurations) and probabilistically (e.g. the FHA CS25 regulatory thresholds). Possibly tuning of the abstraction levels will necessitate the Counter-Example Generative Abstraction Refinement Approach (CEGAR) [Tab10], [Bel17], [Sao18a], [Sao18b]. We intend to explore this control synthesis and verification method uniformly on the digital (software and hardware), the physical, and the governed digital physical.

[70] Predicate abstraction of hybrid system models into state machines may be timed or untimed. Hence this option manifested between parentheses. The timed choice scales less that the untimed one.
[71] FDIR detectors and pass-fail reconfiguration predicates, some qualitative physics predicates (e.g. too fast, too strong, …).

# Conclusion

We have promoted a triple interpretation of cyber in Cyber Physical System: "governed", "digital", and "threat". We have emphasized the role of control in the digital continuum and argued the existence of a *governed*-digital-physical continuum that pervades the digital society.

We have reviewed some of the new behavioral complexity sources that we deem are significant contributors to greater risks, to the widening gap between expected and delivered trust. We have proposed long-term orientations to bridge this gap. It requires to draw from sophisticated mathematics and algorithms on one side, to change the safety engineering mindset on the other side.

We have reviewed a selection of academic results at the confluence of system level formal methods and control engineering. They give us genuine hopes to overcome, in the mid- and long-term, the present pain points. The main ingredients of the proposed research roadmap are set-based engineering, contract-based compositionality, invariant-based design, abstraction-refinement, of the infinite-state to the finite-state compositional and observational reduction, specification model learning, and deterministic-probabilistic co-verification of high fidelity models or development codes.

We have presented the µΩ use case and research project, opened to external collaborations to continue the quest for scalable, affordable and trustworthy engineering of the governed digital-physical. Four years of part-time µ-development has shown that it needs a surprisingly high amount of effort to complete the Ω-development of this µ-CPS/CPSoS. Openness will favor accelerated progress, exciting debates, and dissemination. Sharing of the lessons learnt and co-construction of the successful approaches are a necessary condition for any eligibility in certification as future recommended practices.

# Acknowledgements

I'm grateful to Eric Goubault for his guidance in the bibliographic landscape of hybrid system theory, and for many fruitful discussions and contacts.

My thanks to Ulrich Fahrenberg for useful pointers, comments and supportive exchange while writing this paper.

I'm glad to acknowledge the commitment and contributions to the µΩ project of the members of the System & Software Engineering Laboratory at TRT Palaiseau: Olivier Constant, Eric Dujardin, Florian Greff, Jérôme Le Noir, Sébastien Madelenat, Gilles Malfreyt, Clément Souyri and Thomas Vergnaud.

This work has partially been supported by H2020 ECSEL CPS4EU grant ID 826276.